\documentclass[journal,onecolumn,12pt]{IEEEtran}
\usepackage{cite}
\usepackage{soul,graphicx}
\usepackage[colorlinks,pdfstartview=FitH,citecolor=blue,linkcolor=blue,urlcolor=blue]{hyperref}
\usepackage{amsmath,amssymb,amsthm}
\usepackage{eulervm}
\usepackage{tikz}
\usetikzlibrary{matrix}
\usepackage{mathtools}
\usepackage[braket]{qcircuit}
\newcommand{\op}[2]{|#1 \rangle\langle #2|}
\newcommand{\ip}[2]{\langle #1|#2\rangle}
\newcommand{\ops}[1]{|#1 \rangle\langle #1|}

\DeclareMathOperator{\Tr}{Tr}
\newcommand{\bm}[1]{\mbox {\boldmath $#1$}}
\newcommand{\bms}[1]{\mbox {\boldmath ${}_{#1}$}}
\usepackage[doublespacing,nodisplayskipstretch]{setspace}
\begin{document}
%
\title{\LARGE Quantum Information Processing: An Essential Primer}
%
%
%

\author{Emina Soljanin,~\IEEEmembership{Fellow,~IEEE}
\thanks{E.~Soljanin is with the Department
of Electrical and Computer Engineering, Rutgers University, Piscataway, New Jersey 08854, USA,
e-mail: (see \href{https://www.ece.rutgers.edu/emina-soljanin}{https://www.ece.rutgers.edu/emina-soljanin})}}
%
\markboth{IEEE Journal of on Selected Areas in Information Theory, revised, August~2020}{}
%
%
\maketitle
%
\begin{abstract}
Quantum information science is an exciting, wide, rapidly progressing, cross-disciplinary field, and that very nature makes it both attractive and hard to enter. In this primer, we first provide answers to the three essential questions that any newcomer needs to know: How is quantum information represented? How is quantum information processed? How  is  classical  information  extracted  from  quantum  states?  We then introduce the most basic quantum information theoretic notions concerning  entropy,  sources,  and  channels, as well as secure communications and error correction.  We conclude with examples that illustrate the power of quantum correlations. No prior knowledge of quantum mechanics is assumed.
\end{abstract}



%
\IEEEpeerreviewmaketitle

\section{Introduction\label{sec:intro}}
Quantum phenomena provide computing and information handling paradigms that are distinctly different and arguably much more powerful than their classical counterparts. In the past quarter of the century, much progress has been made on the theoretical side, and experiments have been carried out in which quantum computational operations were executed on a small number of quantum bits (qubits). The US National Science Foundation has declared this general area to be one of the 10 big ideas for future investments. In June 2018, the science committee of the US House of Representatives unanimously approved the National Quantum Initiative Act (H.R.~6227, Public Law No: 115-368) \cite{funding:6227} to create a 10-year federal effort aimed at boosting quantum science. Similar funding commitments have been made throughout the world. 

We have entered what is known as the Noisy Intermediate-Scale Quantum (NISQ) technology era \cite{NISQ:preskill19}. 
This term refers to devices with 50-100 qubits (intermediate-scale), which is  too few to have have full error-correction (noisy).  Nevertheless, NISQ systems  may be able to perform tasks that exceed the capabilities of today's classical digital computers, and may be useful tools for exploring many-body quantum physics.
On the theoretical side, significant progress has been made in understanding the fundamental limits of quantum telecommunications systems, giving rise to the subfield of quantum information theory. Moreover, classical information theory has been used to understand problems in the foundations of physics.

This is an exciting moment to enter the field, and many excellent textbooks and class notes are available to assist you at that (see e.g.\ \cite{QB:mermin,QB:NandC,QB:Wilde,QB:Hidary,QB:QIM}). The goal of the primer is to quickly provide some essential information. It is aimed at a wide audience with diverse background and interests, and assumes no familiarity with quantum mechanics (see e.g. \cite{QB:SandF} for a minimal introduction to the field). It does, however, require undergraduate-level knowledge of probability (see e.g.\ \cite{PB:P-N}) and linear algebra (see e.g.\ \cite{MB:BandF}, Volume 1, or \cite{QB:Hidary}, Part 3).

This primer is organized and can be read as follows.
 Sections~\ref{sec:representing}, \ref{sec:processing}, and \ref{sec:measuring} answer the following three fundamental questions: How is quantum information represented? How is quantum information processed? How is classical information extracted from quantum states? This material is necessary for everyone entering the field regardless of their further interests. Perhaps surprisingly, this material is also sufficient to understand many quantum algorithms, including the celebrated Shor celebrated factoring algorithm (provided you are familiar with their non-quantum background.)  This material is also sufficient to understand the fundamental ideas and principles of quantum key distribution covered in Section~\ref{sec:QKD} and quantum error correction covered in Section \ref{sec:ECC}.
 Section~\ref{sec:mixed} describes a more general framework to represent quantum states, which is needed for quantum information theory as well as for working on quantum computing platforms such as IBM-Q. Section~\ref{sec:IT} introduces the most basic quantum information theoretic notions concerning entropy, sources, and channels. Rudimentary knowledge of classical information theory is necessary for reading this section. Section~\ref{sec:entanglement} illustrates the power of quantum correlations through three canonical examples of ``quantum magic''.
%

\section{How is Quantum Information Represented?}\label{sec:representing}
Representation of quantum information is connected to a postulate of quantum mechanics which says that 
associated to any isolated physical system is a complex vector space  with inner product. In this primer, and quantum computing in general, we mostly deal with finite dimensional spaces $\mathbb{C}^N$ and often conventionally refer to them as Hilbert spaces $\mathcal{H}_N$.

\subsection{Single Qubit}
Classical computers store and operate on bits. What about quantum computers? The quantum information and computing counterpart to the bit is the {\it qubit}.  Qubits (as bits) are represented by physical systems. Mathematically, independently of a particular physical realization, a qubit is represented by a unit-norm\footnote{The norm is induced by the inner product.} vector in the two-dimensional unitary space $\mathbb{C}^2$. If we denote the basis vectors of this space by
\[
{\displaystyle |0\rangle ={\bigl [}{\begin{smallmatrix}1\\0\end{smallmatrix}}{\bigr ]}} 
~~ \text{and} ~~
{\displaystyle |1\rangle ={\bigl [}{\begin{smallmatrix}0\\1\end{smallmatrix}}{\bigr ]}}
\]
then a single qubit $\ket{\psi}$ is mathematically a linear combination of $\ket{0}$ and $\ket{1}$, that is
\begin{equation}
    \displaystyle \ket{\psi} =\alpha \ket{0} +\beta \ket{1}
    \label{eq;qubit}
\end{equation}
where $\alpha$ and $\beta$ are complex numbers such that 
$\displaystyle |\alpha |^{2}+|\beta |^{2}=1$. 
Given this constraint, we can write 
$|\psi \rangle$  uniquely as follows:
\begin{equation}
|\psi \rangle =\cos \left(\theta /2\right)|0\rangle \,+\,e^{i\phi }\sin \left(\theta /2\right)|1\rangle
    \label{eq:psBSr}
\end{equation}
where $0 \leq \theta \leq \pi$ and $0 \leq \phi < 2 \pi$. 
 Expression \eqref{eq:psBSr} can be visualised through the Bloch sphere representation of quantum states, which we introduce in Sec.~\ref{sec:bloch}. 

In classical computing, we refer to a {\it bit value} or a binary value. In quantum computing, we refer to a {\it qubit state} or a quantum state. Rather than a {\it linear combination}, we say that the quantum state $\ket{\psi}$ above is a superposition of the two basis states. The basis $\ket{0}$, $\ket{1}$ in which the qubit is represented is called the {\it computational basis}. The superposition is instrumental in enabling quantum computing parallelism and speedup.

The notation we used for unit-norm column vectors above is common in the quantum computing literature. In this notation, known as the Dirac or bra-ket notation, {\it ket} state $\ket{\psi}$ denotes a unit-norm column vector and {\it bra} state $\bra{\psi}$, its Hermitian transpose (a row vector). The {\it bracket} $\ip{\psi}{\varphi}$ denotes the inner product, hence the name bra-ket. The outer product $\op{\psi}{\varphi}$ is a matrix, and $\ops{\psi}$ is a rank--$1$ projection matrix. The Dirac notation  is preferred by physicists in general, and almost exclusively used. It let us easily tell apart not only scalars and vectors, but also column and row vectors.

\subsection{Multiple Qubits}
Consider two qubits: $\ket{\psi_1}=\alpha_1|0\rangle + \beta_1|1\rangle$ and $\ket{\psi_2}=\alpha_2|0\rangle + \beta_2|1\rangle$.
	 The joint state of the pair  is the Kronecker product of the individual states:
	\begin{align*}
	\ket{\psi_1}\otimes \ket{\psi_2} & = \big(\alpha_1|0\rangle + \beta_1|1\rangle\big)\otimes  \big(\alpha_2|0\rangle + \beta_2|1\rangle\big)\\
	& =
	\alpha_1\alpha_2\ket{0}\otimes\ket{0}+
	\alpha_1\beta_2\ket{0}\otimes\ket{1}+
	\beta_1\alpha_2\ket{1}\otimes\ket{0}+
	\beta_1\beta_2\ket{1}\otimes\ket{1}
	\end{align*}
	where
\[
	\begin{array}{ccccccc} 
	\ket{0}\otimes\ket{0}=\left[\begin{smallmatrix}1\\0\\0\\0\end{smallmatrix}\right]&& 
	\ket{0}\otimes\ket{1}=\left[\begin{smallmatrix}0\\1\\0\\0\end{smallmatrix}\right]&& 
	\ket{1}\otimes\ket{0}=\left[\begin{smallmatrix}0\\0\\1\\0\end{smallmatrix}\right]&& 
	\ket{1}\otimes\ket{1}=\left[\begin{smallmatrix}0\\0\\0\\1\end{smallmatrix}\right] 
	\end{array} 
\]
is a basis for $\mathbb{C}^4=\mathbb{C}^2\otimes\mathbb{C}^2$.	In general, a 2-qubit state is any linear combination of the four basis states, and thus cannot always  be expressed as a Kronecker product of single qubit states. To see this, consider the two-qubit state
$
	|\psi \rangle =\big(|0\rangle \otimes |0\rangle +|1\rangle \otimes |1\rangle\big)/\sqrt{2}.
$
%
States that can be written as a Kronecker product of  single-qubit states are called {\it separable} and those that cannot are called {\it entangled} states.

An $n $--qubit state $\ket{\phi}$ is a unit-norm vector in  $\mathbb{C}^{2^n}$, which we commonly refer to as the Hilbert space ${\cal H}_{2^n}={\cal H}_{2}^{\otimes n}=\underbrace{{\cal H}_2\otimes {\cal H}_2\otimes\dots\otimes {\cal H}_2}_n$.
We will also often use the common notation $N=2^n$.
For an $n $--qubit state  $\ket{\phi}\in {\cal H}_{2^n}$, we have
\[
\ket{\phi} =	\sum_{i=0}^{2^n-1}\alpha_i|i_0i_1\dots i_{n-1}\rangle,\qquad  	\sum_{i=0}^{2^n-1}|\alpha_i|^2=1.
\] 
where $i_0i_1\dots i_{n-1}$ is the binary representation of $i$, and $|i_0i_1\dots i_{n-1}\rangle$ represents shorthand notation for $|i_0\rangle\otimes |i_1\rangle\otimes \dots\otimes  |i_{n-1}\rangle$ (the $i$-th basis vector of ${\cal H}_{2^n}$). Other commonly used shorthand notation is 
\[
|i_0\rangle\otimes |i_1\rangle\otimes \dots\otimes  |i_{n-1}\rangle\equiv
|i_0\rangle|i_1\rangle\dots|i_{n-1}\rangle\equiv
|i_0,i_1,\dots , i_{n-1}\rangle \equiv
|i_0i_1\dots i_{n-1}\rangle
\]
\section{How is  Quantum Information Processed?\label{sec:processing}}	
Processing of quantum information is connected to a postulate of quantum mechanics which says that 
the evolution of a closed quantum system is described by a unitary  transformation. Therefore,
in a closed quantum system, a single-qubit state $|\psi\rangle\in{\cal H}_2$ can be transformed to
some other state in ${\cal H}_2$, say $|\varphi\rangle$, in a reversible way only by some unitary operator $U$, i.e., 
\[
|\varphi\rangle = U|\psi\rangle
\]
where $U$ is a $2\times 2$ unitary matrix over $\mathbb{C}$. Note that quantum evolution is reversible. 

\subsection{Unitary Evolution and Quantum Gates}

Unitary evolution in a closed quantum system is a consequence of the Schr\"{o}dinger equation.
In a closed system,  the state (wave function) $\ket{\psi(t)}$  evolves according to the Schr\"{o}dinger equation:
\[
i\hslash {\frac {d}{dt}}\ket{\psi (t)} = H\cdot \ket{\psi (t)}
\]
where $H$ is a fixed Hermitian matrix known as the system's Hamiltonian.
If $H$ does not depend on time, the solution of this equation is
\[
\ket{\psi(t)} = U(t)\ket{\psi(0)}
\]
and it is easy to show that  $U(t) = \exp\bigl(-\frac{i}{\hslash}Ht\bigr)$ is a unitary matrix.
The Hamiltonian describes the physical model. The Schr\"{o}dinger equation tells us how a state-vector evolves in time given the physical model described by the Hamiltonian.

If we know how $U$ acts on the basis vectors $\ket{0}$ and $\ket{1}$, then we also know how it acts on any vector $\ket{\psi} =\alpha \ket{0} +\beta \ket{1}$ since the evolution (matrix multiplication) is a linear operation, and thus
\[
U\ket{\psi} =\alpha U\ket{0} +\beta U\ket{1}.
\]
Unitary action $U$ maps the computational basis $\ket{0}$, $\ket{1}$ into the basis $U\ket{0}$,  $U\ket{1}$.

	Actions on an $n $--qubit state are described by  $2^n\times 2^n$ unitary matrices, which may or may not 
	 be Kronecker products of matrices of smaller dimensions.
	When $U = U_0\otimes U_1\otimes\dots\otimes U_{n-1}$, where $U_i$ is a $2\times 2$ unitary matrix, then its action on the 
	basis vector 
	$
	|i_0\rangle\otimes |i_1\rangle\otimes \dots\otimes  |i_{n-1}\rangle \in {\cal H}_{2^n}
	$
	is given by
	\[
	U\ket{i_0i_1\dots i_{n-1}}= \boxed{U_0|i_0\rangle}
	\otimes \boxed{U_1|i_1\rangle}\otimes\dots\otimes \boxed{U_{n-1}|i_{n-1}\rangle}
	\]	
	
\subsection{Some Standard Quantum Gates \label{sec:gates}}
	Unitary operators that act on one or more qubits are often referred to as {\it quantum gates}. The requirement that the gates be unitary, i.e., reversible, rules out quantum versions of some classical gates, e.g.,  the {\tt AND} gate but not the {\tt NOT} gate. We next present some often used quantum gates.
\subsubsection{Single Qubit Gates}
In classical computing, \texttt{NOT} is the only  single bit gate, that is, in addition to the {\tt I} ``gate'' (identity). In quantum computing, any $2\times 2$  unitary matrix specifies a single-qubit gate. The most commonly used single-qubit gates are the Pauli and Hadamard matrices.
\\[1ex]
\ul{The Identity and the Pauli Matrices}:\\
The matrices are given below together with their action on the basis vectors in the quantum circuit notation:
\begin{small}
		\begin{center}
				\begin{tikzpicture}
			\node at (0,0) {$I = \begin{bmatrix}
			 1 & 0\\	0 &  1 \end{bmatrix}$};
		 	\node[right] at (0.75,0.35) {\Qcircuit @C=1em @R=.7em {
		 		&&& \lstick{\ket{0}}     & \gate{I} & \rstick{\ket{0}}\qw}};
		 \node[right] at (0.75,-0.35) {\Qcircuit @C=1em @R=.7em {
		 		&&& \lstick{\ket{1}}     & \gate{I} & \rstick{\ket{1}}\qw}};
				\begin{scope}[shift={(0,-1.85)}]
			\node at (-0.15,0) {$\sigma_X = \begin{bmatrix}
				0 &  1 \\ 1 & 0 \end{bmatrix}$};
			\node[right] at (0.75,0.35) {\Qcircuit @C=1em @R=.7em {
					&&& \lstick{\ket{0}}     & \gate{X} & \rstick{\ket{1}}\qw}};
			\node[right] at (0.75,-0.35) {\Qcircuit @C=1em @R=.7em {
					&&& \lstick{\ket{1}}     & \gate{X} & \rstick{\ket{0}}\qw}};
				\end{scope}
				\begin{scope}[shift={(6.5,0)}]
			\node at (-0.15,0) {$\sigma_Y = \begin{bmatrix}
				0 & -i \\ i & 0 \end{bmatrix}$};
		\node[right] at (0.75,0.35) {\Qcircuit @C=1em @R=.7em {
					&&& \lstick{\ket{0}}     & \gate{Y} & \rstick{i\ket{1}}\qw}};
		\node[right] at (0.75,-0.35) {\Qcircuit @C=1em @R=.7em {
					&&& \lstick{\ket{1}}     & \gate{Y} & \rstick{-i\ket{0}}\qw}};
			\end{scope}
			\begin{scope}[shift={(6.5,-1.85)}]
			\node at (-0.15,0) {$\sigma_Z = \begin{bmatrix}
				1 &  0 \\ 0 & -1 \end{bmatrix}$};
		\node[right] at (0.75,0.35) {\Qcircuit @C=1em @R=.7em {
					&&& \lstick{\ket{0}}     & \gate{Z} & \rstick{\ket{0}}\qw}};
		\node[right] at (0.75,-0.35) {\Qcircuit @C=1em @R=.7em {
					&&& \lstick{\ket{1}}     & \gate{Z} & \rstick{-\ket{1}}\qw}};
				\end{scope}
			\end{tikzpicture}
							\vspace{-2ex}
		\end{center}
\end{small}
Note that $\sigma_X$ maps $\ket{0}$ into $\ket{1}$ and vice versa, and thus it is often referred to as the quantum {\tt NOT} or the bit-flip gate.
These four matrices have many interesting properties, and appear often in physics and mathematics. In particular, any $2\times2$ complex matrix  $A$ (and thus any unitary matrix) can be expressed as a linear combination of the identity $I$ and the Pauli matrices $\sigma_X$, $\sigma_Y$, and $\sigma_Z$:
\[
A =\alpha_I I+\alpha_X\sigma_X+\alpha_Y\sigma_Y+\alpha_Z\sigma_Z
\]
for some complex numbers $\alpha_I$, $\alpha_X$, $\alpha_Y$, and $\alpha_Z$. 
\\[0.5ex]
\ul{The Hadamard gate}:
The Hadamard single qubit gate is defined by the normalized Haramard $2\times 2$ matrix. 
		\begin{small}
			\begin{center}
				\begin{tikzpicture}
				\node at (0,0) {$H = \frac{1}{\sqrt{2}}\begin{bmatrix}
					1 &  1 \\ 1 & -1 \end{bmatrix}$};
				\node at (2.4,0.35) {\Qcircuit @C=1em @R=.7em {
						&&& \lstick{\ket{0}}     & \gate{H} & \rstick{(\ket{0}+\ket{1})/\sqrt{2}}\qw}};
				\node at (2.4,-0.35) {\Qcircuit @C=1em @R=.7em {
						&&& \lstick{\ket{1}}     & \gate{H} & \rstick{(\ket{0}-\ket{1})/\sqrt{2}}\qw}};
				\end{tikzpicture}
				\vspace{-2ex}
			\end{center}
		\end{small}
The basis of $\mathbb{C}^2$ defined by vectors 
	\begin{equation}
\ket{+}=(\ket{0}+\ket{1})/\sqrt{2}~~\text{and}~~\ket{-}=(\ket{0}-\ket{1})/\sqrt{2}
\label{eq:HB}
	\end{equation}
 that results from the Hadamard action of the computational basis is known as the Hadamard basis.

\subsubsection{Two Qubit Gates}
The two-qubit quantum gate known as  quantum \texttt{XOR} or controlled-not gate \texttt{CNOT} is specified as a map, a circuit, and a unitary matrix $U_{\texttt{CNOT}}$ as follows:
\begin{center}
	\begin{small}
	\begin{tikzpicture}
	\node at (0,0.35) {
		$\texttt{CNOT}: |x,y\rangle \rightarrow |x,x\oplus y\rangle$};
	\node at (0,-0.35) {
		$x,y\in\{0,1\}$};
	\node at (4,0) {\Qcircuit @C=1em @R=1em {
			\lstick{\ket{x}} & \ctrl{1} & \rstick{\ket{x}} \qw \\
			\lstick{\ket{y}} & \targ & \rstick{\ket{x\oplus y}} \qw
	}};
\node at (8.5,0) {{\small
	$U_{\texttt{CNOT}} =
	\begin{bmatrix}
	1 & 0 & 0 & 0 \\
	0 & 1 & 0 & 0 \\
	0 & 0 & 0 & 1 \\
	0 & 0 & 1 & 0 
	\end{bmatrix}$}
};
	\end{tikzpicture}	
	\end{small}
\end{center}
	We can use the Hadamard and the \texttt{CNOT} gates to create entanglement as follows:
	\begin{center}
		\begin{tikzpicture}
		\node at (0,0) {\Qcircuit @C=1em @R=.7em {
				\ket{0} & & \gate{H} & \ctrl{3} & \qw &\\
				& & & & \rstick{\frac{1}{\sqrt{2}}\bigl(\ket{00}+\ket{11}\bigr)}\\
				& \\
				\ket{0} & & \qw & \targ & \qw &
		}};
	\end{tikzpicture}
	\vspace{-2ex}	
\end{center}
\subsection{The No-Cloning Theorem\label{sec:cloning}}
The requirement that any evolution be unitary gives rise to the famous no-cloning theorem \cite{Cloning:park70,Cloning:dieks82,Cloning:woottersZ82}, which asserts that there is no unitary operator $U_c$ on $\mathcal{H}_{2}\times \mathcal{H}_{2}$ that takes  state $|\psi\rangle\otimes\ket{\omega}$ to
	$|\psi\rangle\otimes |\psi\rangle$ for all states $\ket{\psi}\in \mathcal{H}$ and some fixed state   $\omega\in \mathcal{H}$. 

To prove the no-cloning  theorem, we suppose that there is a unitary matrix $U_c$ such that for two arbitrary sates $\ket{\psi}$ and $\ket{\varphi}$, we have
	\[
	\begin{array}{rcl}
	U_c(|\psi\rangle \otimes |\omega\rangle) & = & |\psi\rangle\otimes |\psi\rangle\\
	U_c(|\varphi\rangle \otimes |\omega\rangle) & = & |\varphi\rangle\otimes |\varphi\rangle
	\end{array}
	\]
	where $\omega$ is some fixed quantum state.
	Note the following identities:
	\begin{enumerate}
		\item  By the properties of the Kronecker product, we have
		\[
		\big(\bra{\psi}\otimes\bra{\omega}\big)\cdot
		\big(\ket{\varphi}\otimes\ket{\omega}\big) = \ip{\psi}{\varphi}
		\]
		\item Since $U_c$ is unitary, that is $U_c^\dag \cdot U_c= I$, then by the properties of the Kronecker product, we have
		\begin{align*}
		\ip{\psi}{\varphi} &=\big(\bra{\psi}\otimes\bra{\omega}\big)\cdot
		\big(\ket{\varphi}\otimes\ket{\omega}\big) \\
		&=\big(\bra{\psi}\otimes\bra{\omega}\big)U_c^\dag \cdot U_c
		\big(\ket{\varphi}\otimes\ket{\omega}\big)\\
		&=\big(\bra{\psi}\otimes\bra{\psi}\big)\cdot
		\big(\ket{\varphi}\otimes\ket{\varphi}\big)\\
		&= \ip{\psi}{\varphi}\otimes \ip{\psi}{\varphi} = \ip{\psi}{\varphi}^2
		\end{align*}
	\end{enumerate}
	Therefore $\ip{\psi}{\varphi}$ is either equal to 0 or to 1.
	Thus if $U_c$ can clone some state $\ket{\psi}$, then the only other state $U_c$ can clone has to be orthogonal to $\ket{\psi}$.

The no-cloning theorem is often misunderstood to be more restrictive than it is. Note that it does not prohibit the following map:
	\[
	\begin{array}{rcl}
	\underbrace{\alpha|0\rangle + \beta|1\rangle}_{\in {\cal H}_2}
	& \rightarrow & \underbrace{\alpha|000\rangle + \beta|111\rangle }_{\in {\cal H}_{2^3}}
	\end{array}
	\]
which can be accomplished by the  circuit consisting of gates we defined above as follows:
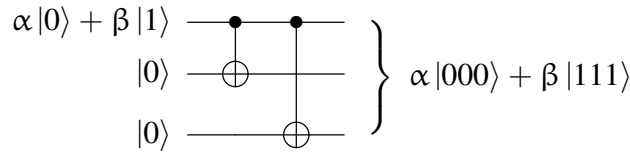
\begin{figure}[hbt]
	\begin{center}
		\begin{tikzpicture}
		\node at (0,0) {\Qcircuit @C=1.1em @R=1.1em {
					\lstick{\alpha\ket{0}+\beta\ket{1}} & \ctrl{1}  & \ctrl{2} &  \qw \\
					\lstick{\ket{0}} & \targ & \qw & \qw\\
					\lstick{\ket{0}} & \qw & \targ & \qw}
		};
		\node at (1.25,0.08) {$\left. \begin{array}{c}
			~\\~\\~
			\end{array}\right \}$};
		\node[right] at (1.75,0.08) {$\alpha\ket{000}+\beta\ket{111}$};
		\end{tikzpicture}
	\end{center}
\caption{Quantum circuit consisting of two {\tt CNOT} gates that maps $(\alpha\ket{0}+\beta\ket{1})\otimes\ket{0}\otimes\ket{0}$ to 
$\alpha|000\rangle + \beta|111\rangle$. }
\label{fig:RepEnc}
\end{figure}

\subsection{Creating Quantum Parallelism}
We can evaluate an $m$-bit valued function $f$ of an $n$-bit string by what is known as the {\it function evaluation gate}. The evaluation gate for a function $f:\{0,1\}^n\rightarrow \{0,1\}^m$ is described as follows:
	\begin{center}
		\begin{tikzpicture}
		\node at (0,0.35) {
			$U_f: |x,y\rangle \rightarrow \ket{x,y\oplus f(x)}$};
		\node at (0.5,-0.35) {
			$x\in\{0,1\}^n$, $y\in\{0,1\}^m$};
		\node at (5,0) {\Qcircuit @C=1em @R=.7em {
				\lstick{\ket{x}} & \multigate{1}{U_f}  & \rstick{\ket{x}}\qw\\
				\lstick{\ket{y}} & \ghost{U^\dag} &\rstick{\ket{y\oplus f(x)}}\qw
		}};
		\end{tikzpicture}
		\vspace{-1ex}
	\end{center}
Note that $U_f$ is a unitary operator acting on vectors in $\mathcal{H}_2^{\otimes n}\otimes \mathcal{H}_2^{\otimes m}$.

We have seen above that the Hadamard gate action on $\ket{0}$ creates a uniform superposition of the computational basis states. It is easy to show that applying the $n$--qubit Hadamard product gate  $H^{\otimes n}$ to $\ket{0}^{\otimes n}$ creates  the uniform superposition of the computational basis of $\mathcal{H}_{2^n}$
	\[
\Qcircuit @C=1em @R=.7em {
	&&& \lstick{\ket{0}^{\otimes n}}     & \gate{H^{\otimes n}} & \rstick{ \frac{1}{{2}^{n/2}}
		{\sum_{x\in\{0,1\}^n}\ket{x}}}\qw}
	\]	
Quantum function evaluation parallelism is achieved by first creating the uniform superposition of the computational basis of $\mathcal{H}_{2^n}$	and then applying the $U_f$ unitary transform to simultaneously evaluate $f$  on its entire domain, as follows:
	\begin{equation}
		U_f\bigl(H^{\otimes n}\otimes I_m\bigr)
		\bigl(\ket{0}^{\otimes n}\otimes\ket{0}^{\otimes m}\bigr)=
		\frac{1}{2^{n/2}}\sum_{x\in\{0,1\}^n}\ket{x,f(x)}.
\label{eq:QP}
\end{equation}
If we could also simultaneously read all the evaluations (which we cannot), we would achieve a quantum speedup. We will see what is possible  in the next section which describes how we can extract classical information from quantum states. 
\section{How is Classical Information Extracted?\label{sec:measuring}}
Extraction of classical  information from quantum states is connected to a postulate of quantum mechanics which says that 
to every physical observable, there corresponds an operator defined by a Hermitian matrix. 
The only possible results of  measuring an observable are the eigenvalues of its corresponding Hermitian matrix. 
The only possible states after  measuring an observable are the normalized (unit-norm) eigenvectors of its Hermitian matrix. 
\subsection{Observables and Expected Values}
\begin{figure}[hbt]
	\centering
	\includegraphics[scale=1.2]{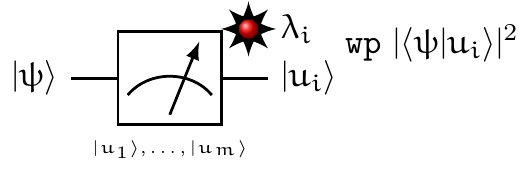}
	\caption{Quantum measurement: The only possible results of  ``measuring $H$'' are its eigenvalues $\lambda_i$, and the only possible states after  ``measuring $H$'' are  its normalized eigenvectors $\ket{u_i}$. When we ``see'' $\lambda_i$
		(which happens wp $|\ip{\psi}{u_i}|^2$ when state $\ket{\psi}$ is measured), we know that state being measured $\ket{\psi}$ has collapsed to $\ket{u_i}$. }
	\label{fig:QM}
\end{figure}

The measurement of an observable $H$ always indicates an eigenvalue of $H$ and turns any measured quantum state  into the eigenstate of $H$ corresponding to the indicated eigenvalue. The measured state only gives rise to a probability distribution on the set of outcomes, as we explain next.
Let $\lambda_1,\dots,\lambda_N$ be the eigenvalues of an $N\times N$ Hermitian matrix $H$ and $\ket{u_1},\dots,\ket{u_N}$ the corresponding  eigenvectors. (We assume, for the moment, that all $\lambda_i$ are different.)
Let $\ket{\psi}$ be a state being measured by the observable described by $H$. Then the measurement result is $\lambda_i$ and  $\ket{\psi}$ collapses to $\ket{u_i}$ with probability (wp) $|\ip{\psi}{u_i}|^2$, $1\le i\le N$, as sketched in Fig.~\ref{fig:QM}.

Since $H$ is hermitian, and thus unitarily diagonizable, it holds that 
\begin{enumerate}
	\item $\ip{u_i}{u_j}=\delta_{ij}$
	\item $\ops{u_1}+\ops{u_2}+\dots+\ops{u_N}=I_N$
\end{enumerate}
A set of vectors $\ket{u_1},\dots,\ket{u_N}$  that satisfies the above two conditions is said to form a \textit{resolution of the identity}. We refer to $\ket{u_1},\dots,\ket{u_N}$ as the measurement basis, and say that we {\it perform a measurement in the basis} or measure in the basis $\ket{u_1},\dots,\ket{u_N}$.

\noindent
{\it Example:} When we measure state $\ket{\psi} =\alpha \ket{0} +\beta \ket{1}$ in the computational basis, the result will be $\ket{0}$ wp $|\alpha|^2$ and $\ket{1}$ wp $|\beta|^2$.
What can we get if we measure qubit $\bigl(\ket{0}+\ket{1}\bigr)/\sqrt{2}$ in the computational basis
$\ket{0}$,  $\ket{1}$? What is the probability of the outcomes? How about the Hadamard basis defined in \eqref{eq:HB}?

Regardless of which state $\ket{\psi}$ is being measured by the observable described by $H$, the only possible outcomes are  the eigenvalues of $H$. The expected value of the measurement depends on $\ket{\psi}$ as follows:
\begin{align*}
\sum_{i=1}^N\lambda_i|\ip{\psi}{u_i}|^2 & =\sum_{i=1}^N\lambda_i\ip{\psi}{u_i}\ip{u_i}{\psi}\\
& = \bra{\psi}\bigl({\textstyle \sum_{i=1}^N\lambda_i\ops{u_i}}\bigr)\ket{\psi} = \bra{\psi}H\ket{\psi}
\end{align*}
where we have used the equality $H=\sum_{i=1}^N\lambda_i\ops{u_i}$. (The reader may care to observe the convenience of the Dirac notation in this simple derivation.)

It is natural to wonder whether these probabilistic measurements can be useful.
Recall that quantum parallelism allows us to evaluate  $f:\{0,1\}^n\rightarrow \{0,1\}^m$ on its entire domain (see map \eqref{eq:QP}).
But we have just seen that we cannot simultaneously extract all the values by a single measurement. How is then quantum speedup achieved? Many quantum algorithms 
prescribe further processing of the state $\frac{1}{2^{n/2}}\sum_{x\in\{0,1\}^n}\ket{x,f(x)}$ so that, when a measurement is eventually performed, the probability of getting the answer to the posed question is (close to) $1$.  Moreover, the questions usually ask about some global property such as whether a function is balanced or constant  or what is its period rather than the explicit function evaluation on its entire domain.

\subsection{Mathematical Description of the Quantum Measurement}
We have seen above that a measurement on an $n$--qubit state is defined by a set of $N=2^n$ basis 
vectors $\ket{u_i}$, $1\le i\le N$. 
When state $\ket{\psi}$ enters the measuring apparatus, it collapses to the state  $\ket{u_i}$ wp $|\ip{\psi}{u_i}|^2$. 
If we denote by $\Pi_i$ the rank--$1$ projection on $\ket{u_i}$, we can equivalently say the the measurement is defined by the set of $N$ orthogonal rank--$1$ projections $\Pi_i=\ops{u_i}$ and the measured state $\ket{\psi}$ collapses to $\frac{1}{\sqrt{\bra{\psi}\Pi_i\ket{\psi}}}\Pi_i\ket{\psi}$ 
wp $\bra{\psi}\Pi_i\ket{\psi}$. We can now easily generalize our basis defined measurement by removing the requirement that the projections $\Pi_i$ be rank--$1$.
\subsubsection{Von Neumann Measurement}
An observable described by a Hermitian $N\times N$ matrix $H$ may have $m\le N$ different eignvalues. Let $\Pi_i$ be the projection operator on the eigenspace $i$ of $H$, $1\le i\le m$.  The von Neumann projective measurement is defined as follows:
		\begin{itemize}
			\item A set of pairwise orthogonal projection operators $\{\Pi_i\}$ such that $\sum_i \Pi_i=I$.
			\item For input $\ket{\psi}$, output $i$ happens with probability $\bra{\psi}\Pi_i\ket{\psi}$, and $\ket{\psi}$ collapses to  $\frac{1}{\sqrt{\bra{\psi}\Pi_i\ket{\psi}}}\Pi_i\ket{\psi}$.
		\end{itemize}
	
\subsubsection{Positive Operator-Valued Measure (POVM)}
We can further generalize the von Neumann measurement of an $n$--qubit state $\ket{\psi}\in\mathcal{H}_{2^n}$ by observing that we can add an ancillary $m$--qubit state in $\mathcal{H}_{2^m}$  to $\ket{\psi}$ and perform  a von Neumann measurement to the joint state in $\mathcal{H}_{2^n}\otimes \mathcal{H}_{2^m}$. If we restrict our attention\footnote{{\it Restricting our attention} to a part of the system is a formal mathematical notion; see Sec.~\ref{sec:bipartite} which discusses bipartite states.} to $\mathcal{H}_{2^n}$, the measurement is defined as follows:
		\begin{itemize}
			\item Any set of  positive-semidefinite operators $\{E_i\}$ such that $\sum_iE_i=I$.
			\item  For input $\ket{\psi}$, output $i$ happens with probability $\bra{\psi}E_i\ket{\psi}$, and $\ket{\psi}$ collapses to  $\frac{1}{\sqrt{\bra{\psi}E_i\ket{\psi}}}\Pi_i\ket{\psi}$.
		\end{itemize}
\subsection{Examples of Quantum Measurements\label{sec:QMexamples}}
We consider measuring two single-qubit states $\ket{\psi_0}$ and $\ket{\psi_1}$. The angle between these vectors is $2\pi/6$.

\subsubsection{Von Neumann Measurement}	
\begin{figure}[hbt]
\begin{center}
	\begin{tikzpicture}
	\node at (0,0) {\includegraphics[scale=1.2]{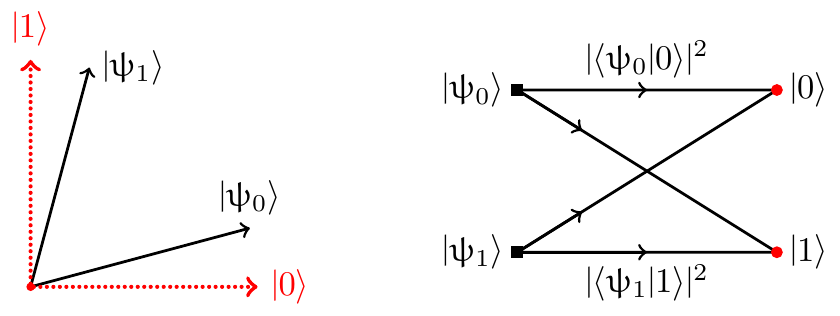}};
	\node at (-5.75,2.0) {(a)};
\node at (5.4,2.0) {(b)};
	\end{tikzpicture}
\end{center}
\caption{(a) States $\ket{\psi_0}$ and $\ket{\psi_1}$ with the bases $\ket{0}$, $\ket{1}$  used for a von Neumann measurement. (b) the states before and after the measurement with the possible transitions. (Some labels are omitted for clarity of the figure.)}
\label{fig:QMbsc}
\end{figure}

The measurement is defined by the computational basis vectors $\ket{0}$ and $\ket{1}$. The angle between $\ket{\psi_0}$ and $\ket{0}$ is $\pi/12$, and so is the angle between $\ket{\psi_1}$ and $\ket{1}$, as in Fig.~\ref{fig:QMbsc}.
No matter which state is measured, the resulting state after the measurement is either  $\ket{0}$ or $\ket{1}$. If state $\ket{\psi_0}$ is measured, it will collapse either to state $\ket{0}$ with probability $|\ip{\psi_0}{0}|^2=\cos^2(\pi/12)=1/2+\sqrt{3}/4$ or to state $\ket{1}$ with probability $|\ip{\psi_0}{1}|^2=1-|\ip{\psi_0}{0}|^2=\sin^2(\pi/12)=1/2-\sqrt{3}/4$.  We can make similar observations when state $\ket{\psi_1}$ is measured.

\subsubsection{Positive Operator Value Measure (POVM)}
\begin{figure}[hbt]
	\begin{center}
		\begin{center}
			\begin{tikzpicture}
			\node at (0,0) {\includegraphics[scale=1.2]{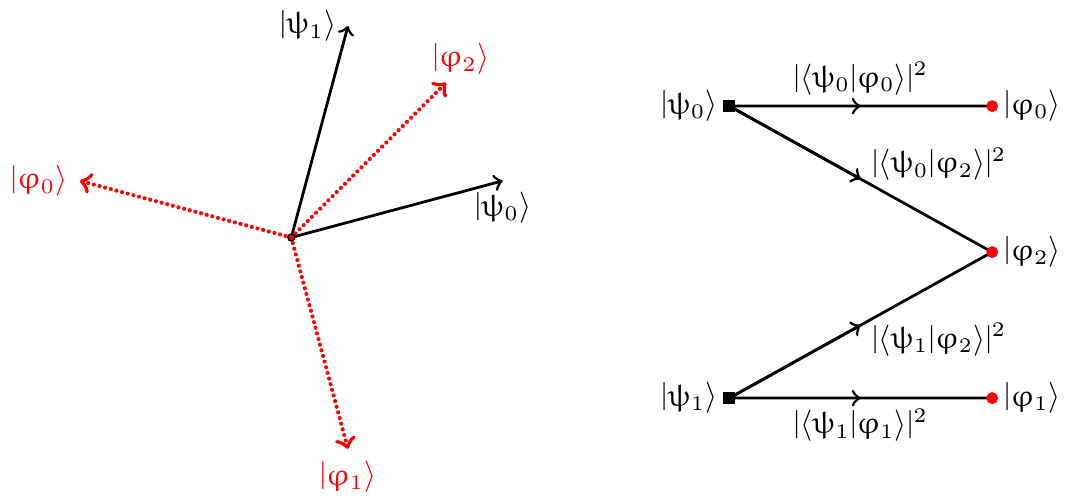}};
			\node at (-6.5,2.75) {(a)};
			\node at (6.5,2.75) {(b)};
			\end{tikzpicture}
		\end{center}
	\end{center}
\caption{(a) States $\ket{\psi_0}$ and $\ket{\psi_1}$ with the vectors $\ket{\varphi_0}$, $\ket{\varphi_1}$, and $\ket{\varphi_1}$ used for a POVM. (b) the states before and after the measurement with the possible transitions.}
\label{fig:QMbec}
\end{figure}

The measurement is defined by the projections on vectors $\ket{\varphi_0}$, $\ket{\varphi_1}$, $\ket{\varphi_2}$. The angle between $\ket{\varphi_0}$ and $\ket{\varphi_i}$, $i=1,2$,  is $2\pi/3$, and it is easy to see that properly normalized projections on these vectors form a resolution of the identity $I_2$.  Vectors $\ket{\psi_0}$ and $\ket{\varphi_1}$  are orthogonal, and so are  $\ket{\varphi_0}$ and $\ket{\psi_1}$, as in Fig.~\ref{fig:QMbec}.
No matter which state is measured, the resulting state after the measurement is one of the states   $\ket{\varphi_0}$, $\ket{\varphi_1}$, $\ket{\varphi_2}$. 
If state $\ket{\psi_0}$ is measured, it will collapse either to state $\ket{\varphi_0}$ with probability $|\ip{\psi_0}{\varphi_0}|^2=\cos^2(2\pi/6)=1/4$ or to state $\ket{\varphi_2}$ with probability $|\ip{\psi_0}{\varphi_2}|^2=\cos^2(\pi/6)=3/4$. 
Note that the probability of state $\ket{\psi_0}$ collapsing to $\ket{\varphi_1}$ is zero. We can make similar observations when state $\ket{\psi_1}$ is measured.

\subsubsection{Measurements Defined by Pauli Matrices}
Pauli Matrices are both unitary and Hermitian, and thus can serve to define both quantum gates and quantum  measurements. Their	eigenvalues with the corresponding eigenvectors are shown in Table~\ref{tab:PauliEV}.
\begin{table}[hbt]
		\caption{Pauli matrices and their eigenvalues with the corresponding normalized eigenvectors.}
	\begin{center}
\begin{tabular}{c|llll}
	matrix &&\multicolumn{3}{c}{eigenvalue/eigenvector} \\[1pt]
	\hline\\[-7pt]
	$\sigma_X$ && $+1/(|0\rangle+|1\rangle)$ && $-1/(|0\rangle-|1\rangle)$\\[1ex]
	$\sigma_Y$ &&  $+1/|0\rangle+i|1\rangle)$ &&  $-1/(|0\rangle-i|1\rangle)$\\[1ex]
	$\sigma_Z$  &&  $+1/|0\rangle$ && $-1/|1\rangle$
\end{tabular}
\end{center}
\label{tab:PauliEV}
\end{table}

\section{Quantum Key Distribution\label{sec:QKD}}
Traditional data encryption methods, based on using public keys, are threatened by the advances in quantum computing algorithms promising to efficiently solve so far intractable problems that make public key encryption currently secure.  However, it is precisely quantum information processing advances that are also expected to enable secure communications by allowing efficient and secure private key distribution. The main advantage of private key encryption is that as long as the key strings are truly secret, it is  provably secure, that is, insensitive to advances in computing. 

A Quantum Key Distribution (QKD) protocol describes how two parties, commonly referred to as Alice and Bob, can establish a secret key by communicating over a quantum and a public classical channel when both channels can be accessed by an eavesdropper Eve. The basic observation behind QKD protocols is that, since Eve cannot clone qubits (cf.\ Sec.~\ref{sec:cloning}), she can only gain information by measuring the original qubit.
Therefore, when non-orthogonal qubits are transmitted from Alice to Bob, then Eve cannot gain any information from the qubits without disturbing their states, thus alerting Alice and Bob of her presence. We next describe two important QKD protocols. Substantial progress has been made towards building practical schemes based on these protocols. 

		\subsection{BB84 Protocol} 
		The BB84 was developed by Bennett and Brassard in 1984, hence the name \cite{QKD:BennettB84}. We outline steps that
		Alice and Bob make under this protocol in order to generate a secret key of $O(n)$ bits for an arbitrary integer $n$.
		
		\begin{enumerate}
			\item Alice creates a sequence of  $(4 + \delta)n$ random data bits which she will map into qubits for transmission over the quantum channel between her and Bob. 
			\item For each data bit, Alice tosses a fair coin. If she gets a head (H), she maps her data bit into either $\ket{0}$ (if her data bit is $0$) or $\ket{1}$ (if her data bit is $1$). If she gets a tail (T), she maps her data bit into either $\ket{+}$ (if her data bit is $0$) or $\ket{-}$ (if her data bit is 
			$1$). Recall that $\ket{-}$ and $\ket{+}$ states are defined in \eqref{eq:HB}.
			We will refer to the sequence of heads and tails that Alice generated  as $C_A$. We will call $\{\ket{0},\ket{1}\}$ the $H$ basis and $\{\ket{-},\ket{+}\}$ the $T$ basis, according to the corresponding coin faces.
			\item  Alice sends the resulting $(4 + \delta)n$ qubits to Bob over their public quantum communication channel. Each qubit may be altered by the noise in the channel and/or measured by Eve. Note that, at this point, Eve has no knowledge of $C_A$
				and thus  what measurement basis she should use for an intercepted qubit 
				in order to learn the corresponding bit. She can only guess the  preparation basis  for a qubit, and if her guess is wrong, she will alter its state, thus leaving a proof of eavesdropping.
			
			\item Upon receiving a qubit, Bob   then tosses a fair coin and then, depending on the toss outcome, he measures the qubit in either the $H$ or the $T$ basis. If he uses the $H$ bases and gets $\ket{0}$, or the $T$ bases and gets $\ket{+}$, he records bit 0; otherwise he records bit 1. 
			\\[0.75ex]
			We refer to the sequence of heads and tails  generated by Bob as $C_B$.
			\item Once Bob receives $(4 + \delta)n$ qubits, Alice publicly announces $C_A$  and  Bob publicly announces $C_B$. 
			\item Alice and Bob discard the bits where sequences $C_A$ and $C_B$ differ (that is, when Bob measured a qubit a in the different basis than Alice used for its preparation). With high probability, there are at least 2n bits left (if not, repeat the protocol). They keep $2n$ bits.
			\item Alice selects a subset of $n$ bits from the remaining $2n$ that will serve to check Eve's interference, and tells Bob which bits she selected. 
			\item Alice and Bob announce and compare the values of the $n$ check bits. If more than an acceptable number disagree, they abort the protocol. (The acceptable number is determined by e.g., the noise in the channels.)
			\item Alice and Bob perform classical information reconciliation and privacy amplification on the remaining $n$ bits to obtain  $O(n)$ shared key bits.
		\end{enumerate}
		
		\subsection{E91-like Protocols}
		The E91 protocol was proposed by Ekert in 1991, hence the name \cite{QKD:ekert91}. The scheme distributes entangled pairs of photons so that Alice and Bob each end up with one photon from each entangled pair. The creation and distribution of photons can be done by Alice, by Bob, or by some third party. 
		
		Suppose Alice and Bob share a set of $n$ entangled pairs of qubits in the state $(\ket{00}+\ket{11})/\sqrt{2}$
		and Eve is not present. If they measure their respective states in the computational basis, they will get identical sequences of completely random bits. Thus the scheme benefits from two properties of shared entanglement: randomness and correlation.  
		To check if Eve was present, Alice and Bob can, for example, select a random subset of the shared entangled pairs, and test to see if they are entangled (instead of using them to generate the key bits). They can do that e.g., by playing the CHSH game discussed in Sec.~\ref{sec:CHSH}.

		\section{Error Correcting Codes\label{sec:ECC}}
Error correcting codes add redundancy to data in order to make it less sensitive to errors. The most basic form of redundancy is simple replication (cloning), known as {\it repetition coding}. For example, if each bit is replicated 3 times, any single bit flip among the 3 replicas can be corrected by turning it to the value of the other two replicas, after first finding out (measuring) what the value of the majority is. But could there be a counterpart to this process in the quantum world where the no-cloning theorem holds and the measurements disturb the states?\footnote{{\it As significant as Shor's factoring algorithm may prove to be, there is another recently discovered feature of quantum information that may be just as important: the discovery of quantum error correction. Indeed, were it not for this development, the prospects for quantum computing technology would not seem bright.}
	\\
	\strut\hfill John~Preskill,
	{\it Quantum Computation Lecture Notes.} Chapter~1, 1997/98.}
We will first formally describe the process of introducing redundancy (encoding) and correcting errors (decoding) for a 1-to-3 bits repetition code, which will allow us to introduce and understand its quantum 1-to-3 qubit counterpart.
\subsection{A Classical Error Correcting Code}
\begin{itemize}
\item \ul{Encoding} is a map that introduces redundancy. In our  1-to-3 bits repetition code example, each bit $x$ is mapped to a 3 bit string $x\, x\, x$, that is, the encoding is the following map:
\[
0\rightarrow 000 ~~\text{and} ~~ 1 \rightarrow 111
\]
\item \ul{Error Model:} In this example, at most one of the bits $x\,x\,x$ gets flipped. Such flipping is equivalent to adding (component-wise) a string in the set $\{000,100,010,001\}$ to 
$x\,x\,x$ and getting $y_0\,y_1\,y_2$:
\begin{center}
	\begin{tabular}{c||ccc}
		additive error & $y_0$ & $y_1$ &$y_2$\\
		\hline\hline
		$0\,0\,0$ & $x$ &$x$ & $x$ \\
		$1\,0\,0$ & $x\!\oplus\!1$ &$x$&$x$\\
		$0\,1\,0$ & $x$& $x\!\oplus\!1$ &$x$\\
		$0\,0\,1$ & $x$&$x$& $x\!\oplus\!1$
	\end{tabular}
\end{center}
\item \ul{Measurements:} We perform the following matrix vector multiplication (cf.\ two measurements):
\begin{equation}
\begin{bmatrix}
1 & 1 & 0\\
1 & 0 & 1
\end{bmatrix}
\cdot
\begin{bmatrix}
y_0\\y_1\\y_2
\end{bmatrix}
= 
\begin{bmatrix}
y_0\oplus y_1\\y_0\oplus y_2
\end{bmatrix}
\label{eq:syndrome}
\end{equation}
We refer to the vector $\begin{bmatrix}
y_0\oplus y_1\\y_0\oplus y_2
\end{bmatrix}$ as the {\it error syndrome}. Observe that the first bit of the syndrome tells us if whether bits $y_0$ and $y_1$ have identical values and  the second bit of the syndrome tells us if whether bits $y_0$ and $y_2$ have identical values.
\vspace{1ex}
\item \ul{Error Correction:} The 2-bit measurement result (syndrome) tells us which bit is flipped, and thus instructs us how to correct errors:
\begin{center}
\begin{tabular}{ccc||c|c||c}
$y_0$&$y_1$&$y_2$ &$y_0\oplus y_1$ & $y_0\oplus y_2$
 & add to $y_0\,y_1\,y_2$\\
 \hline\hline
x&x&x & $0$ & $0$ & $0\,0\,0$\\
$x\oplus$ 1&x&x & $1$ & $1$ & $1\,0\,0$\\
x& $x\oplus$ 1&x & $1$ & $0$ & $0\,1\,0$\\
x&x&$x\oplus 1$ & $0$ & $1$ & $0\,0\,1$\\
\end{tabular}
\end{center}
\end{itemize}

\subsection{A Quantum Error Correcting Code}
Quantum error correction has to follow the laws of quantum mechanics. Therefore all actions on qubits (encoding, errors, decoding) have to be either unitary or measurements.  We describe the simplest code only to show that quantum error correction under these constraints is feasible, and possibly make the reader interested in this fascinating subject.
\footnote{{\it Correcting errors might sound like a dreary practical problem, of little aesthetic or conceptual interest. But aside from being of crucial importance for the feasibility of quantum computation, it is also one of the most beautiful and surprising parts of the subject. }\\
\strut\hfill David~Mermin,
{\it Quantum Computer Science: An Introduction.} Cambridge Univ.\ Press.}
\begin{itemize} 
\item \ul{Encoding:} As in the classical case, encoding is a map that introduces redundancy. In our example, a single qubit state is mapped into a 3-Qubit state as follows:
\[
\alpha\ket{0}+\beta\ket{1}\rightarrow \alpha\ket{000}+\beta\ket{111}
\]
The unitary circuit shown in Fig.~\ref{fig:RepEnc} can serve as a quantum mechanically valid encoder for our code. It uses two {\tt CNOT} gates and two ancillary qubits, each initially in the state $\ket{0}$. The result is an entangled 3-Qubit state.

\item 
\ul{Error Model:} We assume that at most one qubit experiences the basis flip (i.e., is acted on by $\sigma_X$, see Sec.~\ref{sec:gates}). The possible 3-qubit error operators and the resulting states they give when acting on $\alpha\ket{000}+\beta\ket{111}$ are as follows:
\begin{center}
\begin{tabular}{c||c}
error operators & resulting state \\
\hline\hline
$I\otimes I\otimes I$ & $\alpha|000\rangle + \beta|111\rangle$ \\
$\sigma_X\otimes I\otimes I$ & $\alpha|100\rangle + \beta|011\rangle$ \\
$I\otimes \sigma_X\otimes I$ & $\alpha|010\rangle + \beta|101\rangle$ \\
$I\otimes I\otimes \sigma_X$ & $\alpha|001\rangle + \beta|110\rangle$ \\
\end{tabular}
\vspace{1ex}
\end{center}
\item \ul{Measurements:} 
As in the classical case, the idea is to have two measurements such that one compares qubits 1 and 2, and the other compares qubits 1 and 3. The additional constraint here is that the measuring process leave the measured states unchanged. 
We perform the following two measurements: \\[1ex]
$\boxed{M_1}:$ This measurement is defined by the Hermitian operator 
$\sigma_Z\otimes \sigma_Z\otimes I$, i.e.,
the following two orthogonal projection operators:
\begin{align*}
\Pi_1 &=  \ops{000}+\ops{111}
+\ops{001}+\ops{110}\\
\Pi_2 &= \ops{010}+\ops{101}
+\ops{011}+\ops{100}
\end{align*}
$\Pi_1$ projects on the eigenspace of $\sigma_Z\otimes \sigma_Z\otimes I$ with eigenvalue 1, and $\Pi_2$ projects on the eigenspace of $\sigma_Z\otimes \sigma_Z\otimes I$ with eigenvalue $-1$. (See Table~\ref{tab:PauliEV}.)
\\[3ex]
$\boxed{M_2}:$ This measurement is defined by the Hermitian operator 
$\sigma_Z\otimes I\otimes \sigma_Z$, i.e.,
the following two orthogonal projection operators:
\begin{align*}
\Pi_1 &=  \ops{000}+\ops{111}
+\ops{010}+\ops{101}\\
\Pi_2 &= \ops{001}+\ops{110}
+\ops{011}+\ops{100}
\end{align*}
$\Pi_1$ projects on the eigenspace of $\sigma_Z\otimes I\otimes \sigma_Z$ with eigenvalue 1, and $\Pi_2$ projects on the eigenspace of $\sigma_Z\otimes I\otimes \sigma_Z$ with eigenvalue $-1$. (See Table~\ref{tab:PauliEV}.)

\item \ul{Error Correction:} The results of the two measurements are two eigenvalues (2 bits). As in the classical case, we refer to this result as the error syndrome, which  instructs us how to correct errors, as follows:
\begin{center}
\begin{tabular}{c||c|c||c}
corrupted state & $M_1$
& $M_2$ & apply \\
\hline\hline
$\alpha|000\rangle + \beta|111\rangle$ & $+1$ & $+1$ & $I\otimes I\otimes I$\\
$\alpha|100\rangle + \beta|011\rangle$ & $-1$ & $-1$ & $\sigma_X\otimes I\otimes I$\\
$\alpha|010\rangle + \beta|101\rangle$ & $-1$ & $+1$ & $I\otimes \sigma_X\otimes I$\\
$\alpha|001\rangle + \beta|110\rangle$ & $+1$ & $-1$ & $I\otimes I\otimes \sigma_X$
\end{tabular}
\end{center}
\end{itemize}
\vspace{1ex}
{\it Remark:}  The error detecting procedure we used 
1) follows directly from classical error correction and 2) it is is useful in generalizing to other quantum codes with more qubits. However, $M_1$ and $M_2$ are not the only measurements we can use to obtain the error syndrome that can uniquely identify the error. To see that consider the following set of projections:
\begin{align*}
\Pi_1 = \ops{000}+\ops{111} & ~~ \text{no error}\\
\Pi_2 = \ops{100}+\ops{011} & ~~ \text{bit flip on Qubit one}\\
\Pi_3 = \ops{010}+\ops{101} & ~~ \text{bit flip on Qubit two}\\
\Pi_4 = \ops{001}+\ops{110} & ~~ \text{bit flip on Qubit three} 
\end{align*}
Note that the (no)-error states belong to orthogonal subspaces, and therefore a von Neumann measurement defined by projectors to those subspaces can 1) unambiguously identify the error state and 2) will not disturb the measured state. 

As  their classical counterparts, decoders of quantum error correcting codes can miss-correct or not-detect certain error patterns. For example, the decoder above will miss-correct the two-qubit error  introduced by the operator $\sigma_X\otimes\sigma_X\otimes I$, and it will not detect  three-qubit error $\sigma_X\otimes\sigma_X\otimes\sigma_X$ and even a single-qubit  error $\sigma_Z\otimes I\otimes I$.

\section{Mixed States\label{sec:mixed}}
There are many scenarios when we do not know the state of a quantum system, but do know that it is in the state $|\psi _{j}\rangle$ with probability $p_j$, $j=1,\dots,k$. We say then that the quantum system is in a {\it mixed state}, and refer to the collection of pairs $\{\psi_j,p_j\}_{j=1}^k$ as an {\it ensemble of states}. The states we worked with so far, which can be described by a vector, are known as {\it pure states}.
A mixed state arises e.g., when we know that a measurement of a pure state has been performed but do not know the outcome, or when one of some $k$ noise operators acted on a pure state, as in Sec.~\ref{sec:ECC}. We will use the mixed state notion extensively in the next section when we present some elements of quantum information theory.
This section is concerned with representing, processing, and measuring mixed states.

\subsection{The Density Matrix Formalism}
So far, we used unit-norm vectors in Hilbert spaces to mathematically specify quantum states. We can instead describe a quantum state, say $\ket{\psi}$, by the projection matrix $\rho_{\psi}=\ops{\psi}$.
We refer to $\rho_{\psi}$ as the {\it density matrix} of  $\ket{\psi}$. 
To see that this is a valid model, we next describe 1) how a state evolves when a unitary transformation is applied to it and 2) what happens to a state and with what probability when a measurement is performed on it. 
\begin{enumerate}
	\item Suppose that unitary operator $U$ acts on state $\ket{\psi}$ giving the state $\ket{\varphi}=U\ket{\psi}$.
	We have $\ops{\varphi}=U\ops{\psi}U^{\dag}$. Therefore, 
	$\ket{\psi}\xrightarrow{U} U\ket{\psi}$ is replaced by 
$\rho_{\psi}\xrightarrow{U} U\rho_{\psi}U^{\dag}$.
	\item Suppose a measurement defined by the basis $\ket{u_1},\dots,\ket{u_N}$  is performed on the state $\ket{\psi}$.
	We know that the resulting state will be  $\ket{u_i}$ with probability  (wp) $|\ip{\psi}{u_i}|^2$, $1\le i\le N$.
	In terms of density matrices, we have 
	$
	\rho_{\psi}  \rightarrow  \ops{u_i} ~~ \text{wp} ~~ \Tr \bigl( \ops{u_i} \rho_{\psi}\bigr),
	$
	where the probability expression was obtained by observing that
	$
	|\ip{\psi}{u_i}|^2=\ip{\psi}{u_i}\ip{u_i}{\psi}=\Tr \bigl(\ops{u_i}\cdot\ops{\psi}\bigr)=\Tr \bigl(\ops{u_i}\cdot \rho_{\psi}\bigr).
	$
\end{enumerate}

Note that to describe the state, evolution, and measurement, we used density matrices rather than state vectors.
The advantage of the density matrix formalism is that it allows us to compactly describe mixed states. A mixed state, that is, a quantum system about which we only know that it is in the state $|\psi _{j}\rangle$ with probability $p_j$  has a density matrix defined as follows:
\begin{equation}
\rho =\sum _{j}p_{j}|\psi _{j}\rangle \langle \psi _{j}|.
\label{eq:MSdef}
\end{equation}

The states that have  rank--$1$ density matrices $\rho_{\psi}=\ops{\psi}$ are known as {\it pure states}. 
In general, a density matrix $\rho$ is a Hermitian, positive semi-definite, trace--$1$ matrix. These properties easily follow from \eqref{eq:MSdef}. 

\begin{figure}[hbt]
	\centering
	\includegraphics{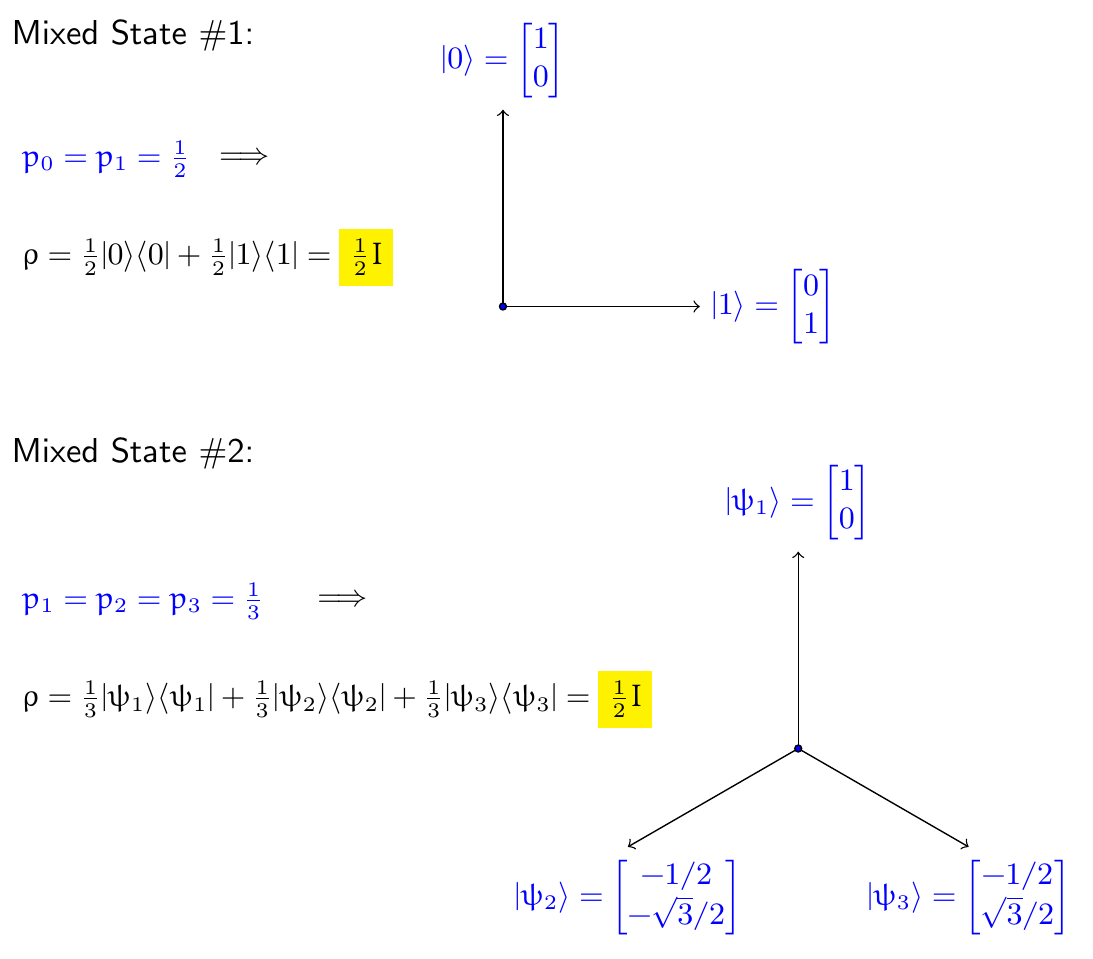}
	\caption{Two ``different'' mixtures of pure states with identical density matrices.}
	\label{fig:2mixtures}
\end{figure}

Observe that two different ensembles of states can have identical density matrices, and therefore quantum mechanically represent identical states. Fig.~\ref{fig:2mixtures} shows two different ensembles with the density matrix equal to $\frac{1}{2}I$.
When a $d\times d$ density matrix is equal to $\frac{1}{2}I$, we say that the system is in the maximally mixed state. These density matrices are quantum counterparts to classical uniform  distributions.

\subsection{Unitary Evolution of Mixed States}
What happens to a mixed state when a unitary transform $U$ is applied to it? If the system described by the mixed state is actually in pure state $\ket{\psi_j}$ with the density matrix $\rho_j=\ops{\psi_j}$, then it will evolve to the state $U\rho_jU^{\dag}$, as we showed above. But we only know that the system is in the state $\rho_{j}$ with probability $p_j$. Therefore, the mixed state will evolve to the state  $U\rho_{j}U^{\dag}$ with probability $p_j$. Therefore, the system with density matrix \eqref{eq:MSdef} will evolve into another mixed state, whose density matrix is given by
\[
\sum _{j}p_{j}U|\psi _{j}\rangle \langle \psi _{j}|U^{\dag}
= U\Bigl(\sum _{j}p_{j}|\psi _{j}\rangle \langle \psi _{j}|\Bigr)U^{\dag} = U\rho U^{\dag} 
\]
Therefore, $\rho\xrightarrow{U} U\rho U^{\dag}$.


\subsection{Measuring Mixed States}
We next look into what happens when we perform a quantum measurement defined by operators $\Pi_i$ on a mixed state whose density matrix is
$\rho =\sum _{j}p_{j}|\psi _{j}\rangle \langle \psi _{j}|$.
Again, let $\rho_j=\ops{\psi_j}$.
If the state being measured is $\ket{\psi_j}$ (which happens with probability $p_j$), then the probability of getting measurement result $i$ is $\Tr \bigl(\Pi_i\rho_j\bigr)$.
Therefore, by the total probability formula, when measuring $\rho$, we get outcome $i$ with probability
\[
\sum_{j}p_{j}\underbrace{\Tr \bigl(\Pi_i\ops{\psi_j}\bigr)}_{\Pr(i|j)} =
\Tr \bigl( \Pi_i\sum _{j}p_{j}\ops{\psi_j}\bigr)
=\Tr \bigl(\Pi_i\rho\bigr)
\]
	Note that different ensembles $\{\psi_j,p_j\}$ with the same $\rho$ will give outcome $i$ with the same probability $\Tr \bigl(\Pi_i\rho\bigr)$, which depends only on $\rho$.

Is the state corresponding to outcome $i$ pure or mixed?  If the state being measured is $\ket{\psi_j}$ and the measurement result is $i$, then the system is in the state $\frac{\Pi_i\rho_{j}\Pi_i}{\Tr \bigl(\Pi_i\rho_j\bigr)} $. Therefore, if we observe outcome $i$,
the system is in the mixed state
\[
\sum_{j}p_{j}\frac{\Pi_i\rho_j\Pi_i}{\Tr \bigl(\Pi_i\rho_{j}\bigr)} =\frac{\Pi_i\rho\Pi_i}{\Tr \bigl(\Pi_i\rho\bigr)}.
\]
Note that we ended up having a mixed state after the measurement resulted in outcome $i$, because we started with a mixed state. 

Which state would we have if we lost the measurement record?  Note that, mathematically,  the state of the system is for us described based on our {\it ignorance/knowledge}. We saw that we get state $\frac{\Pi_i\rho\Pi_i}{\Tr \bigl(\Pi_i\rho\bigr)}$ wp
$\Tr \bigl(\Pi_i\rho\bigr)$. If we lost the measurement record, we would  have a state described by the density matrix
\begin{equation}
\sum_{i=1}Tr \bigl(\Pi_i\rho\bigr)\cdot \frac{\Pi_i\rho\Pi_i}{\Tr \bigl(\Pi_i\rho\bigr)} =\sum_{i=1}\Pi_i\rho\Pi_i.
\label{eq:MSM}
\end{equation}
 Elementary probability  (e.g., the total probability expression) is used for derivations.

\subsection{Bipartite  States\label{sec:bipartite}}
	
Let ${\cal H}_{A}$ and ${\cal H}_{B}$ be finite-dimensional Hilbert spaces with basis states 
$\{|{a_{i}}\rangle \}_{{i=1}}^{n}$ and 
$\{|{b_{j}}\rangle \}_{{j=1}}^{m}$, respectively. 
Then the state space of the composite system is the tensor product ${\cal H}_{A}\otimes {\cal H}_{B}$ with the basis 
$\{|{a_{i}}\rangle \otimes |{b_{j}}\rangle \}$, or in more compact notation 
$\{|a_{i}b_{j}\rangle \}$, $i=1,\dots,n$, $j=1,\dots,m$.

Let $|\psi \rangle \in {\cal H}_{A}\otimes {\cal H}_{B}$. We say that $\ket{\psi}$ is a bipartite pure state of a composite  system  with  subsystems $A$ and $B$. (The state space of a composite physical system is the tensor product of the state spaces of the component physical systems.)
Any pure state of the composite system can be written as
\[ |\psi \rangle =\sum _{i=1}^n \sum _{j=1}^m
	c_{i,j}(|a_{i}\rangle \otimes |b_{j}\rangle )=\sum _{i,j}c_{i,j}|a_{i}b_{j}\rangle , \quad \sum_{i,j} |c_{i,j}|^2 = 1.
\]
When $|\psi \rangle \in {\cal H}_{A}\otimes {\cal H}_{B}$ can be written in the form 
$|\psi \rangle =|\psi _{A}\rangle \otimes |\psi _{B}\rangle$, we say that subsystems $A$ and $B$ are separable. Otherwise they are entangled. When a system is in an entangled pure state, it is not possible to assign states to its subsystems. 

Let $\rho_{AB}$ be a density matrix in the product Hilbert space 
${\cal H}_A\otimes {\cal H}_B$.
A mixed state of the bipartite system described by $\rho_{AB}$ can be
\begin{enumerate}
	\item  a product state if $\rho_{AB} = \rho_A\otimes\rho_B$,
	\item a separable state if there exist a probability distribution 
	$\{p_{k}\}$ and mixed states
	$\{\rho _{A}^{k}\}$ and $\{\rho _{B}^{k}\}$ (states of the respective subsystems) such that
	$\rho =\sum _{k}p_{k}\rho _{A}^{k}\otimes \rho _{B}^{k}$,
	\item an entangled state if neither 1) nor 2) holds.  
\end{enumerate}
Thus for mixed states, {\it separable} and {\it product} are different notions.

Recall the trace expression in the Dirac notation:
Let $\ket{e_i}$, $i=1,\dots, n$ be an orthonormal basis  of $\mathbb{C}^n$. Then $\ops{e_i}A$ is a matrix whose $i$-th diagonal element is $a_{ii}$ and all other diagonal elements are 0. Therefore,
\[
\Tr{A}=\sum_{i=1}^n\Tr\bigl({\ops{e_i}A}\bigr)
=\sum_{i=1}^n\bra{e_i}A\ket{e_i}.
\]

\subsection{Reduced Density Matrices\label{sec:PartialTrace}}
Let $\rho_{AB}$ be a density matrix in the product Hilbert space 
${\cal H}_A\otimes {\cal H}_B$,
and  $\ket{b_i}$ the basis of $ {\cal H}_B$.
The \textit{reduced density matrix} $\rho_A$ is obtained by taking the \textit{partial trace} of $\rho_{AB}$
over the Hilbert space ${\cal H}_B$ as follows:
\[
\rho_{A} = \Tr{}_B\rho_{AB} =
\sum_i \bigl(I\otimes \bra{b_i}\bigr)\rho_{AB}
\bigl(I\otimes \ket{b_i}\bigr).
\]
(You will often see a shorthand expression $ \Tr{}_B\rho_{AB} = \sum_b\bra{b}\rho_{AB}\ket{b}$.)
We say that $\rho_{A}$ is a reduced density operator obtained from $\rho_{AB}$ by {\it tracing out} the subsystem $B$.  Note that the reduced density operators are quantum counterparts to marginal distributions of the classical world. 
\\[1ex]
{\it Example \#1 -- Product State:}\\
Suppose a quantum system is in the product state $\rho_{AB} = \rho_A\otimes\rho_B$ where $\rho_{A}$ is a density operator for system A, and $\rho_{B}$ is a density operator for system B. Then 
\[
\rho_{A} = \Tr{}_B(\rho_A\otimes\rho_B) = \rho_A\Tr\rho_B=\rho_A.
\]
{\it Example \#2 -- Entangled State:}\\
Consider the bipartite state $\ket{\phi_{AB}}=(\ket{00} + \ket{11})/\sqrt{2}$. This is a pure state with the density operator 
\[
\rho_{AB} = \ops{\phi_{AB}} = \frac{1}{2}(\ops{00}+\op{00}{11}+\op{11}{00}+\ops{11}).
\]
Tracing out the second qubit, we find the reduced density operator of the first qubit,
\begin{align*}
\rho_{A} =  \Tr{}_B\rho_{AB} & =(I\otimes \bra{0})\rho_{AB}(I\otimes \ket{0}) + (I\otimes \bra{1})\rho_{AB}(I\otimes \ket{1})\\
& =\frac{1}{2}(\ops{0}+\ops{1})=\frac{1}{2}I.
\end{align*}

Suppose we have $n$--qubit state $\rho$  and an ancillary $m$--qubit state (which can be used to describe interactions with the environment). If we apply a unitary evolution to the joint bipartite state and then trace out the ancillary subsystem, we get that $\rho$ has been transformed to another state $\mathcal{E}(\rho)$ by what is known as a {\it completely positive, trace-preserving} map:
\begin{equation}
\mathcal{E}(\rho) = \sum_kE_k\rho E_k^\dag ~~ \text{where} ~~ \sum_kE_k^\dag E_k=I.
\label{eq:QE}
\end{equation}
Similarly, if we perform a von Neumann measurement on the joint state, and then trace out the ancillary subsystem, we observe that this process is mathematically equivalent to performing a POVM on the original state (cf.\ Sec.~\ref{sec:measuring}).

\subsection{Bloch Sphere\label{sec:bloch}}
The Bloch sphere provides a useful way to represent and visualize both pure and mixed states, and is traditionally used in quantum mechanics. It is also used in quantum computing platforms, such as IBM-Q,  since actions of single-qubit gates on pure states are easy to see within the Bloch sphere  framework.

Any $2\times2$ complex matrix, and thus any density matrix $\rho$, can be expressed as a linear combination of the identity $I$ and the Pauli matrices $\sigma_X$, $\sigma_Y$, and $\sigma_Z$:
\[
\rho =\alpha_I I+\alpha_X\sigma_X+\alpha_Y\sigma_Y+\alpha_Z\sigma_Z
\]
for some complex numbers $\alpha_I$, $\alpha_X$, $\alpha_Y$, and $\alpha_Z$. Since a density matrix is Hermitian and has trace one, these numbers will satisfy certain constraints.

Note that $\sigma_X$, $\sigma_Y$, and $\sigma_Z$  have trace equal to 0. Therefore
\[
\rho ={\frac {1}{2}}\left(I+\beta_X\sigma_X+\beta_Y\sigma_Y+\beta_Z\sigma_Z\right)
\]
where $\beta_X$, $\beta_Y$, and $\beta_Z$  are real numbers.  To see that we write the above expression for $\rho$ as follows:
\begin{equation}
\rho ={\frac {1}{2}} \begin{bmatrix}
1+\beta_Z & \beta_X-i\beta_y\\
\beta_X+i\beta_y & 1-\beta_Z
\end{bmatrix}.
\label{eq:BlochMatrix}
\end{equation}
We call $\vec{\beta}=(\beta_X, \beta_Y, \beta_Z)$  the Bloch vector of $\rho$.
Since $\rho$ is positive semi-definite, we have $\det(\rho)\ge 0$:
\[
0\le \det(\rho) = 1- (\beta_X^2+\beta_Y^2 +\beta_Z^2) = 1-|\vec{\beta}|^2,
\]
which implies $|\vec{\beta}|^2\le 1$. The set of all vectors that satisfy this condition is a ball in $\mathbb{R}^3$, known as the {\it Bloch sphere.}
\begin{figure}[hbt]
	\begin{center}
		\includegraphics[scale=0.9]{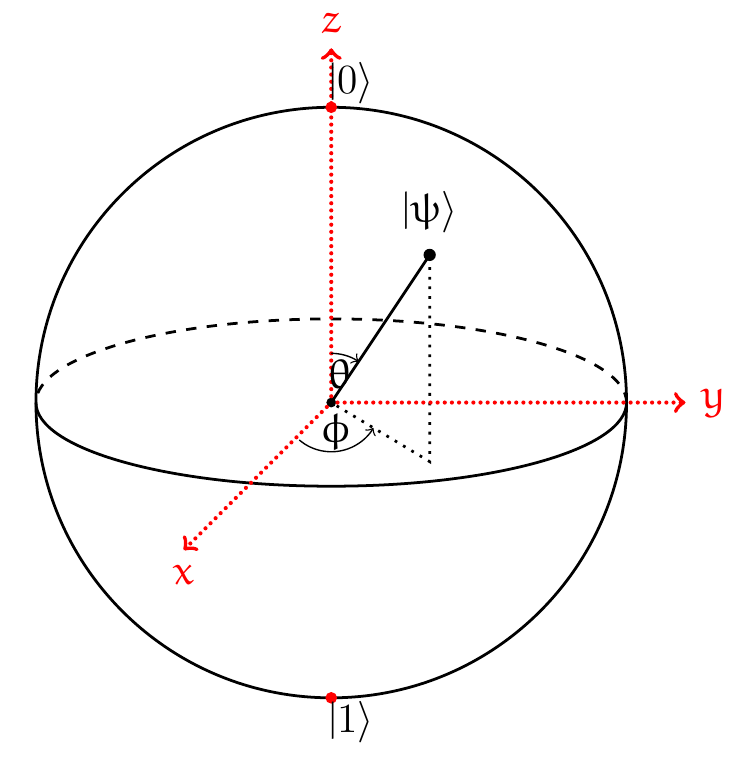}
	\end{center}
	\caption{Bloch Sphere: Pure states $|\psi \rangle =\cos \left(\theta /2\right)|0\rangle \,+\,e^{i\phi }\sin \left(\theta /2\right)|1\rangle$ 
		correspond to the points on the surface, and  mixed states correspond to the points in the interior.}
	\label{fig:bloch_sphere}
\end{figure}

For pure states, we have 
$\mathrm {tr} (\rho ^{2})=1$, and thus
\[
1= \mathrm {tr} (\rho ^{2})={\frac {1}{2}}\bigl(1+|{\vec {\beta}}|^{2}\bigr)\quad \Leftrightarrow \quad |{\vec {\beta}}|=1
\]
Therefore, the surface of the Bloch sphere represents all the pure states of a two-dimensional quantum system, whereas the interior corresponds to all the mixed states. 

We can also see that pure states are points on the Bloch sphere by considering the representation of $\ket{\psi}$ as in \eqref{eq:psBSr}. Comparing $\ops{\psi}$ with the matrix \eqref{eq:BlochMatrix}, we find that the Bloch vector of $\ket{\psi}$ makes an angle of $\theta$ with the $z$ axis, and its projection in the $x-y$ plane makes an angle of $\phi$ with the $x$ axis, as shown in Fig.~\ref{fig:bloch_sphere}.
With the representation \eqref{eq:psBSr}, it is easy to see that
any two diametrically opposite (antipodal) points correspond to a pair of mutually orthogonal pure state vectors. 
In particular, $\beta_X=\beta_Y=0$ and $\beta_Z=1$ gives $\rho=\ops{0}$, while $\beta_X=\beta_Y=0$ and $\beta_Z=-1$ gives $\rho=\ops{1}$.

\section{Elements of Quantum Information Theory\label{sec:IT}}
This section covers the most basic notions of quantum information theory. It defines quantum (von Neumann) entropy, two measures of similarity between quantum states, and the fundamentals concerning quantum sources and channels.
\subsection{Von Neumann Entropy\label{sec:VNE}}

Shannon entropy measures the expected uncertainty associated with a classical probability distribution. In classical information theory, Shannon entropy has multiple {\it operational} meanings, e.g., the compression rate of classical DMS information sources. The quantum counterpart of a probability distribution is a density matrix $\rho$.  Recall that a density matrix $\rho$ is a Hermitian, positive semi-definite, trace one matrix. It follows that $\rho$ can be diagonalized by a unitary matrix, and has eigenvalues that are all real, nonnegative, and sum to one.

The von Neumann  entropy is an older concept that generalizes the Shannon entropy. It is defined as
\[
S(\rho) =  -\Tr\rho\log\rho.
\]
If $\lambda_i$ are the eigenvalues of $\rho$, we have
\[S(\rho)= - \sum_i \lambda_i \log \lambda_i.
\]
The von Neumann  entropy of a density matrix is, therefore, the Shannon entropy of the set of its eigenvalues.

We next list several important properties of the von Neumann  entropy, which we will use in the following sections. Except for \ref{item:aVNE} and \ref{item:ssaVNE}, they easily follow from the definition of von Neumann entropy and the corresponding  properties of Shannon entropy.
\begin{enumerate}
	\item $S(\rho)\ge 0$ with equality iff $\rho$   is a pure state.
	\item $S(\rho) \le \log d$ with equality iff the $\rho=I/d$.
	\item $S(\rho)$ is invariant under a change of the basis.
\item \label{item:aVNE} $S(\rho _{AB}) \le S(\rho _{A})+S(\rho _{B})$ with equality iff $\rho_{AB}=\rho _{A}\otimes \rho _{B}$ \textit{(additivity)}
	\item \label{item:ssaVNE} $S(\rho _{ABC})+S(\rho _{B})\leq S(\rho _{AB})+S(\rho _{BC})$ \textit{(strong subadditivity)}
\end{enumerate}
Proving the strong subadditivity property is considered a significant achievement in quantum information theory \cite{EntropyProp:liebR73}.
Unlike Shannon entropy, von Neumann entropy of a bipartite system may be smaller than that of its subsystems, that is, we cannot claim that $S(\rho_{AB})\ge S(\rho_{A})$ always holds. Consider, for example, the 2-qubit state $\ket{\phi_{AB}}=(\ket{00} + \ket{11})/\sqrt{2}$. This is a pure state  and thus its von Neumann entropy is $0$. However, each of its single-qubit subsystems has  density matrix $\frac{1}{2}I$ (see Sec.~\ref{sec:PartialTrace}), and thus each has von Neumann entropy equal to $1$.

\subsection{Fidelity and Trace Distance} 
Claude Shannon has written that the {\it fundamental problem of communication is that of reproducing at one point either exactly or approximately a message selected at another point}. To  say what {\it approximately} means in quantum communications, we need to have a notion of distance and/or similarity between quantum states.

To measure how faithfully mixed state $\sigma$ approximates mixed state $\omega$ and vice versa, 
we use the so called {\it mixed state fidelity} $F$ defined as 
\[
F(\sigma,\omega) = \Bigl\{\Tr\bigl[(\sqrt{\sigma}\omega\sqrt{\sigma})^{1/2}\bigr]\Bigr\}^2, 
\] 
Besides computing the mixed state fidelity, we can 
measure how close state $\sigma$ is to state $\omega$ by computing the 
{\it trace distance} 
\[ 
D(\sigma,\omega) = \frac{1}{2}\Tr|\sigma-\omega|. 
\] 
where $|A|$ is the positive square root of $A^{\dagger} A$, {\it i.e.,} 
$|A|=\sqrt{A^{\dagger} A}$.  The trace distance is a metric on the space of density operators, and is
closely related to the fidelity as follows:
\begin{equation} 
1-F(\sigma,\omega)\le D(\sigma,\omega)\le \sqrt{1-F(\sigma,\omega)^2}. 
\label{eq:ftd} 
\end{equation} 

The trace distance and the fidelity generalize the classical measures of distance/similarity between probability distributions. 
When matrices $\sigma$ and $\omega$ are simultaneously diagonalizable, the trace distance is equal to the total variation  between their eigenvalues, and the fidelity is equal to the squared  Bhattacharyya coefficient of the their eigenvalues.

\subsection{Compressing Quantum Discrete Memoryless Sources }
A discrete memoryless source (DMS) of information produces a 
sequence of independent, identically distributed random variables 
taking values in a finite set called the {\it source alphabet} ${\cal X}$. 
The source produces letter $a\in {\cal X}$ with probability $p_a$. 
In quantum systems, source letters are mapped into {\it quantum 
	states} for quantum transmission or storage.  In this section, we outline a way to compress pure state sources where source letter $a\in {\cal X}$  is mapped into qubit $\ket{\psi_a}$.  Generalization to higher dimensional spaces is straightforward. While  compression of pure state sources has been known for a quarter of the century \cite{SourceCoding:Schumacher95}, compression of mixed state sources has not been fully understood yet.
A DMS of qubits is completely specified by the ensemble $\mathcal{E}=\{\ket{\psi_a},p_a\}_{a\in \mathcal{X}}$. We refer to the density matrix of this ensemble $\rho = \sum_{a\in{\cal X}} p(a)\ops{\psi_a}$ as the source density matrix.
Note that we can express $\rho$ in terms of its eigenvectors and eigenvalues as follows:
\[ 
	\rho = \lambda_0\ops{\phi_0} +  \lambda_1\ops{\phi_1},
\] 
where $\{\lambda_0, \lambda_1\}$ is a probability mass function
(PMF) on $\{0,1\}$ and $\ip{\phi_0}{\phi_1}=0$. Therefore, $\rho$ is also the density matrix of the ensemble $\{\ket{\phi_i},\lambda_i\}_{i\in \{0,1\}}$, and we
can express $\rho^{\otimes n}$ in two ways as follows:
\[
\rho^{\otimes n} =
\sum_{\bm{x}\in {\cal X}^n}p_{\bms{x}}\ops{\Psi\bms{x}}
= \sum_{\bm{z}\in \{0,1\}^n}\lambda_{\bms{z}}\ops{\Phi\bms{z}}.
\]
where
$\ket{\Psi\bms{x}}
=\ket{\psi_{x_1}}\otimes\dots \otimes \ket{\psi_{x_n}}$,
$p\bms{x}=p_{x_1}\cdot\ldots\cdot p_{x_n}$, $x_i\in {\cal X}$,\\
and
$\ket{\Phi\bms{z}}
=\ket{\psi_{z_1}}\otimes\dots \otimes \ket{\psi_{z_n}}$,
$\lambda\bms{z}=\lambda_{z_1}\cdot\ldots\cdot \lambda_{z_n}$, $z_i\in \{0, 1\}$.

We can define typical sequences according to the distribution $\Lambda=\{\lambda_0,\lambda_1\}$ as follows.
We say that sequence $\bm{z} = z_1,\dots,z_n\in\{0,1\}^n$ is weakly $\epsilon_n$-typical  if
\begin{equation}
2^{-n(S(\rho)+\epsilon)}\leqslant \lambda\bms{z}\leqslant 2^{-n(S(\rho)-\epsilon)}.
\label{eq:Tprob}   
\end{equation}
The set of all such sequences $A_{\epsilon}$  is the typical set according to distribution $\Lambda$. It's size $\bigl|{A_\varepsilon }\bigr|$ is bounded as follows:
\begin{equation}
(1-\varepsilon )2^{n(S(\rho)-\varepsilon )}\leq
\bigl|A_\epsilon \bigr|\leqslant 2^{n(S(\rho)+\epsilon )}.
\label{eq:Tsize}   
\end{equation}
From \eqref{eq:Tprob} and \eqref{eq:Tsize}, it follows that the probability of the typical set $A_{\epsilon}$ is greater than $1-\epsilon$, and thus the probability of its complement $A_{\epsilon}^C$ is 
\begin{equation}
\text{Pr}\bigl(A_{\epsilon}^C\bigr) = \sum_{\lambda_{\bms{z}} \in A_{\epsilon}^C}
\lambda_{\bms{z}} \le \epsilon.
\label{eq:Tps}
\end{equation}

We define the \textit{typical subspace} $\bm{\Lambda_n}$  to be  the subspace spanned by the typical states $\ket{\Phi\bms{z}}$, 
$\bm{z}\in A_{\epsilon}$. 
\begin{figure}[hbt]
	\centering
	\includegraphics{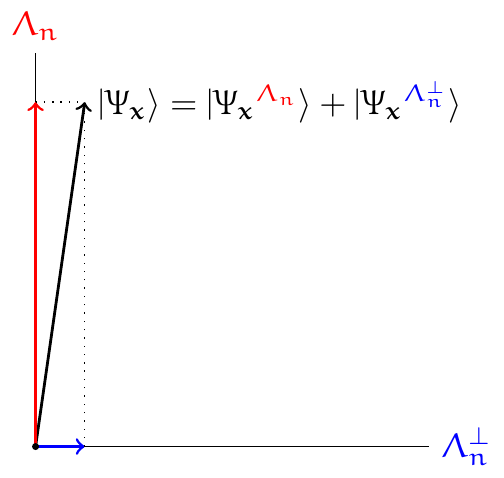}
	\caption{Each source vector state $\ket{\Psi\bms{x}}$ is a sum of its projections to the typical subspace $\bm{\Lambda}$  and its complement $\bm{\Lambda}^\perp$. A source vector is on average well approximated by its projection on the typical space. }
	\label{fig:TS}
\end{figure}
We define the projector on $\bm{\Lambda_n}$ 
and its complement as follows:\\[1ex]
$\Pi=\sum_{\bm{z}\in {A_{\epsilon}}}\ops{\Phi\bms{z}}$
is the projector on $\bm{\Lambda_n}$,
and
$\Pi^\perp=\sum_{\bm{z}\in \{0,1\}^n\setminus{A_{\epsilon}}} \ops{\Phi\bms{z}}$
is the projector on $\bm{\Lambda_n}^\perp$. 
\\[1ex]
Note that   $\Pi+\Pi^\perp=I_{2^n}$.
The dimension of $\bm{\Lambda_n}$ is at most
$2^{n(S(\rho)+\epsilon)}$, cf.~\eqref{eq:Tsize}.

We show that the  fidelity between state $\ket{\Psi\bms{x}}$ and its projection $\Pi\ket{\Psi\bms{x}}$ on the typical subspace is on average high, as follows:
\begin{align*}
\bar{F} & = \sum_{\bm{x}\in {\cal X}^n} P(\bm{x})
F(\ket{\Psi\bms{x}},\Pi\ket{\Psi\bms{x}}) \\
& = \sum_{\bm{x}\in {\cal X}^n}P(\bm{x}) |\bra{\Psi\bms{x}}\Pi\ket{\Psi\bms{x}}|^2\ge
\sum_{\bm{x}\in {\cal X}^n}P(\bm{x}) (1-2\bra{\Psi\bms{x}}\Pi\ket{\Psi\bms{x}})\\
& =  -1 +2\Tr(\Pi\rho^{\otimes n}) \\ 
& = -1 +2\Tr \Bigl\{ \Bigl[ \sum_{\bm{z}\in {A^\Lambda_{\epsilon_n}}} \ops{\Phi\bms{z}}\Bigl]\cdot 
\Bigl[\sum_{\bm{z}\in  \{0,1\}^n} \lambda(\bm{z})\ops{\Phi\bms{z}}\Bigl]\Bigl\}\\ 
& = 1-2\epsilon_n 
\end{align*} 
Therefore, if an $n $--qubit source state $\ket{\Psi\bms{x}}$ is measured by the projectors $\Pi$, $\Pi^\perp$, the outcome will likely collapse to a vector in the typical subspace  $\Pi\ket{\Psi\bms{x}}$ that on average closely approximates $\ket{\Psi\bms{x}}$, as sketched in Fig.~\ref{fig:TS}.
\subsection{Accessible Information \label{sec:acc}}
The accessible information is defined for an ensemble of quantum states as the maximal number of bits that can be communicated by the  ensemble source Alice to the receiver Bob under all quantum measurements that Bob can make. The accessible information cannot exceed a quantity known as the Holevo information of the ensemble for Alexander Holevo who proved this result. In this section, we define the accessible information, state the Holevo bound and outline its proof. 

\subsubsection{Holevo Bound Statement and Implications}
Suppose a source Alice has a classical random variable $X$ over alphabet $\mathcal{X}$ with $|\mathcal{X}|$ letters and letter probabilities $\{p_1, p_2, \dots, p_{|\mathcal{X}|}\}$. When $X$ assumes letter $a\in \mathcal{X}$, Alice prepares a quantum state with density matrix $\rho_a$ and gives this state to the receiver Bob, whose goal is to find the value $a$ of $X$ that Alice has. In order to achieve this goal, Bob performs a measurement on the received state obtaining a classical outcome, namely, a random variable which we call $Y$.

The process of measurement is mathematically equivalent to single-letter classical information transmission through a classical channel. The channel  (that is, its output alphabet and the input-output transition probabilities) is determined by the selected measurements. To see that, consider the ensemble of two pure single-qubit states $\ket{\psi_0}$ and $\ket{\psi_1}$ as in Sec.~\ref{sec:QMexamples}. Fig.~\ref{fig:QMbsc} shows a von Neumann measurement and the associated binary symmetric channel with the crossover probability  $p=\sin^2(\pi/12)=1/2-\sqrt{3}/4$.  Fig.~\ref{fig:QMbec} shows a POVM and the associated binary erasure channel with the erasure probability $\epsilon=3/4$. Thus, for the same input, each measurement selection defines a different channel.

The Alice's RV $X$ is specified by the associated quantum ensemble $\mathcal{E}=\{\rho_a,p_a\}_{a\in \mathcal{X}}$, and Bob's RV $Y$ is specified by $\mathcal{E}$  and the measurement. The maximum value of the mutual information $I(X;Y)$ between the random variables $X$ and $Y$ over all the possible measurements that Bob can make on $\mathcal{E}$ is in quantum information theory known as  the \textit{accessible information} of the ensemble $\mathcal{E}$, and denoted by $\text{Acc}(\mathcal{E})$. We have
\[
\text{Acc}(\mathcal{E}) =\max_{\text{measurements}} I(X;Y)
\]

The general formula to compute the accessible information for an ensemble $\mathcal{E}=\{\rho_a,p_a\}_{a\in \mathcal{X}}$ 
is not known. The best known upper bound on accessible information is the famous Holevo bound \cite{HB:holevo73}:
\begin{equation}
\text{Acc}(\mathcal{E})\leq S(\rho)-\sum _{a\in\mathcal{X}}
p_{a}S(\rho _{a})
\label{eq:HBound}
\end{equation}
where $\rho=\sum _{a\in\mathcal{X}}p_{a}\rho _{a}$ is the ensemble density matrix.
The quantity 
\[
\chi= S(\rho)-\sum _{a\in\mathcal{X}}p_{a}S(\rho _{a})
\]
is called the Holevo information or the Holevo $\chi$ quantity. Note that for an ensemble of pure states, we have $\chi= S(\rho)$.

Consider again the example with two single-qubit states $\ket{\psi_0}$ and $\ket{\psi_1}$ as in Sec.~\ref{sec:QMexamples}, and assume that their probabilities are $P_X(0)=P_X(1)=0.5$. It is an easy but useful exercise to show that $S(\rho) = 0.8112781$.
Fig.~\ref{fig:QMbsc} shows a von Neumann measurement and the induced binary symmetric channel with the crossover probability  $p=1/2-\sqrt{3}/4$. The mutual information is computed as $I(X,Y)=1-h(p)=0.6454211$, where $h(p)$ is the binary entropy. Fig.~\ref{fig:QMbec} shows a POVM and the induced binary erasure channel with the erasure probability $\epsilon=1/4$. The mutual information  is computed as $I(X,Y)= 1-\epsilon = 1/4$.

The Holevo bound, in particular, tells us that the most Alice can communicate to Bob by single qubit is a single bit of classical information, regardless of how large her ensemble of states is (cf.\ Sec.~\ref{sec:acc}).  Note that her ensemble can be arbitrarily large since a qubit, say $\ket{\psi}=\alpha\ket{0}+\beta\ket{1}$, is specified by two complex numbers $\alpha$ and $\beta$. However, Bob cannot read the values of complex numbers, that is {\it inaccessible to him}. He can only possibly apply some unitary transformations to $\ket{\psi}$ and then perform a measurement, which would give him at most one bit. 

\subsubsection{Holevo Bound Proof Outline}
In order prove the Holevo bound \eqref{eq:HBound}, we devise a tripartite quantum system we refer to $ABM$. The subsystem $A$ corresponds to the Alice's RV $X$, and is described by the ensemble $\{\ket{a},p_a\}_{a\in \mathcal{X}}$ where $\ket{a}$, ${a\in \mathcal{X}}$ are orthogonal states. The subsystem $B$ corresponds to the quantum state prepared by Alice and given to Bob and is described by the ensemble $\mathcal{E}=\{\rho_a,p_a\}_{a\in \mathcal{X}}$. The subsystem $M$ (channel output RV $Y$) is where Bob imprints his  measurement result.

The joint system $ABM$ is prior to the measurement in the state
\[
\rho_{{ABM}}=\sum _{a\in\mathcal{X}}p_{a}|a\rangle \langle a|\otimes \rho _{a}\otimes |0\rangle \langle 0|
\]
The state of the subsystem $M$ before the measurement is some known state of the register, here $\ops{0}$.
If the $B$ subsystem is in state $\rho_a$ (for some $a\in\mathcal{X}$, and Bob performs the von Neumann measurement $\{\Pi_m\}$, then Bob is left with the state (cf.~\eqref{eq:MSM}):
\[
\rho _{a}\otimes \ops{0} \longrightarrow \sum_m \Pi_m\rho_a\Pi_m \otimes \ops{m}
\]
Therefore, the tripartite system state $\rho_{{ABM}}$ is mapped into $\rho_{{ABM}}^\prime$ as follows:
\[
\rho_{{ABM}}\longrightarrow \rho_{{ABM}}^\prime = 
\sum _{a\in\mathcal{X}}p_{a}|a\rangle \langle a|\otimes \sum_m \Pi_m\rho_a\Pi_m \otimes \ops{m}
\]
By the strong subadditibvity of the quantum entropy (see Sec.~\ref{sec:VNE}), we have
\begin{equation}
S(\rho_{{ABM}}^\prime) + S(\rho_M^\prime)\le S(\rho_{{AM}}^\prime)+S(\rho_{{BM}}^\prime).
\label{eq:ss}
\end{equation}
The Holevo bound follows from \eqref{eq:ss} and the following, easy to prove, identities:
	\begin{enumerate}
		\item $S(\rho_{{ABM}}^\prime)= H(X) + \sum _{a\in\mathcal{X}}p_{a}S(\rho _{a})$
		\item $S(\rho_M^\prime) = H(Y)$
		\item $S(\rho_{{AM}}^\prime) = H(X,Y)$
		\item $S(\rho_{{BM}}^\prime) = S(\rho)$
	\end{enumerate}

\section{Entanglement and Quantum Correlations\label{sec:entanglement}}
Entangled states (those that cannot be written as a Kronecker product of  single-qubit states) are responsible for much of ``quantum magic'', and the phenomenon of entanglement is considered to be a cornerstone of quantum computing. This section presents three examples that illustrates the power of entanglement. 

\subsection{Bell States (aka EPR Pairs)}
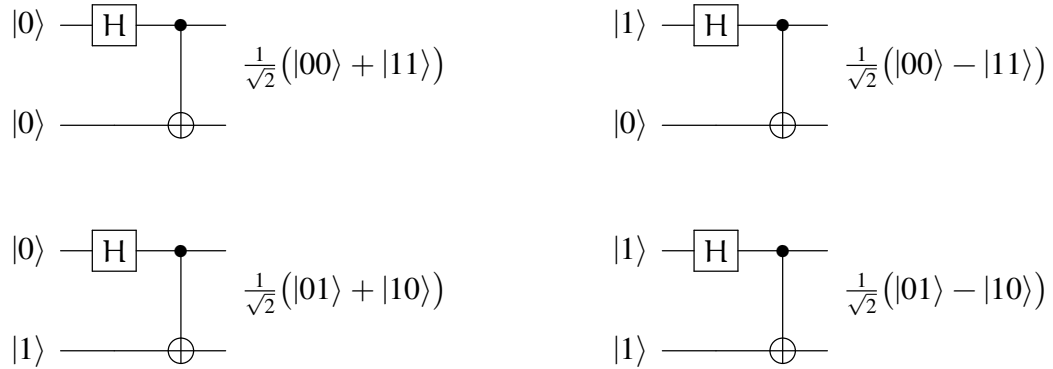
\begin{figure}[hbt]
	\begin{center}
	\begin{tikzpicture}
	\node at (0,0) {\Qcircuit @C=1em @R=.7em {
			\ket{0} & & \gate{H} & \ctrl{3} & \qw &\\
			& & & & \rstick{\frac{1}{\sqrt{2}}\bigl(\ket{00}+\ket{11}\bigr)}\\
			& \\
			\ket{0} & & \qw & \targ & \qw &
	}};
	\node at (0,-3) {\Qcircuit @C=1em @R=.7em {
			\ket{0} & & \gate{H} & \ctrl{3} & \qw &\\
			& & & & \rstick{\frac{1}{\sqrt{2}}\bigl(\ket{01}+\ket{10}\bigr)}\\
			& \\
			\ket{1} & & \qw & \targ & \qw &
	}};
	\node at (8,0) {\Qcircuit @C=1em @R=.7em {
			\ket{1} & & \gate{H} & \ctrl{3} & \qw &\\
			& & & & \rstick{\frac{1}{\sqrt{2}}\bigl(\ket{00}-\ket{11}\bigr)}\\
			& \\
			\ket{0} & & \qw & \targ & \qw &
	}};
	\node at (8,-3) {\Qcircuit @C=1em @R=.7em {
			\ket{1} & & \gate{H} & \ctrl{3} & \qw &\\
			& & & & \rstick{\frac{1}{\sqrt{2}}\bigl(\ket{01}-\ket{10}\bigr)}\\
			& \\
			\ket{1} & & \qw & \targ & \qw &
	}};
	\end{tikzpicture}
		\end{center}
	\caption{Creating Bell states by a 2-qubit entanglement gate.}
	\label{fig:BellStatesCircuit}
\end{figure}

  An entangled pair of states can be created by applying a unitary transform to separable states, e.g., as shown in Fig.~\ref{fig:BellStatesCircuit}. 
These four entangled states are known as  Bell states  or EPR pairs. EPR stands for Einstein, Podolsky and Rosen, who were the first to point out the ``strange'' properties of this state \cite{EPR:einstein35}.  




Notice that the Bell states form a basis (known as Bell's basis), which should not be a surprise since they are created by a unitary transform from the four computational basis states. Therefore, Bell states can be used to define a measurement, which is often referred to as the Bell measurement.

Entangled states have some ``surprising'' properties. To see that, we consider the state
$\bigl(\ket{00}+\ket{11}\bigr)/\sqrt{2}$ and observe the following:
\begin{enumerate}
	\item The individual qubits that make up an entangled state cannot  be characterized as having individual states of their own.  Consider, for example, the first qubit, and observe that it cannot be represented in the form $\alpha\ket{0}+\beta\ket{1}$.
	\item There seems to be {\it spooky action at a distance:}\footnote{Einstein's phrase; he was not comfortable with the notion of non-deterministic measurements and entanglement.} 
	What happens if we measure only the first qubit in the computational basis? Two outcomes are possible: $\ket{0}$ with probability 1/2, giving the post-measurement 2-qubit state $\ket{00}$, and $\ket{1}$ with probability 1/2, giving  the post-measurement 2-qubit state $\ket{11}$. What happens if we subsequently measure the other qubit? Only one outcome is possible: the one that gives the same result as the measurement of the first qubit. This behavior has been confirmed by experiment. 
\end{enumerate}
\subsection{Dense Coding}
The most Alice can communicate to Bob by sending him a single qubit is a single bit of information regardless of how large her ensemble of states is (cf.~Sec.~\ref{sec:acc}). That is,  unless they share an EPR pair. Then a quantum communication protocol known as {\it dense coding} enables Alice  to transmit two classical bits of information (00, 01, 10 or 11) to Bob, by sending him only one qubit \cite{DenseCoding:bennettW92}. Here is how.

Suppose Alice and Bob had prepared together an entangled pair of qubits in the state 
\begin{equation}
\ket{\varphi_{AB}}=\frac{1}{\sqrt{2}}\bigl(\ket{0_A0_B}+\ket{1_A1_B}\bigr)
\label{eq:EPRpair}
\end{equation}
and then Alice took qubit $A$ and Bob took qubit $B$. How does the state $\ket{\varphi_{AB}}$ evolve if only Alice applies a unitary transformation to her qubit? Consider the following four local unitary actions on the first qubit:
\begin{align*}
(I\otimes I) \ket{\varphi_{AB}}& = \frac{1}{\sqrt{2}}\bigl(\ket{0_A0_B}+\ket{1_A1_B}\bigr)\\
(\sigma_X\otimes I)\ket{\varphi_{AB}} & = \frac{1}{\sqrt{2}}\bigl(\ket{1_A0_B}+\ket{0_A1_B}\bigr)\\
(\sigma_Z\otimes I)\ket{\varphi_{AB}} & = \frac{1}{\sqrt{2}}\bigl(\ket{0_A0_B}-\ket{1_A1_B}\bigr)\\
(\sigma_Z\sigma_X\otimes I) \ket{\varphi_{AB}} & = \frac{1}{\sqrt{2}}\bigl(-\ket{1_A0_B}+\ket{0_A1_B}\bigr)
\end{align*}

Note that Alice is able to create four orthogonal states (which would be impossible to do by local actions if the qubits were not entangled). If after performing local action, Alice sends her qubit to Bob, he can unambiguously identify which of the four orthogonal Bell states the EPR pair assumed as a result of Alice's action, by performing a measurement in the Bell basis. He can therefore get two bits of information.  Alice and Bob have to have agreed on  how to label Alice's actions, e.g.,
\begin{align*}
00: & \qquad (I\otimes I)\\
01: & \qquad (\sigma_X\otimes I)\\
10: & \qquad (\sigma_Z\otimes I)\\
11: & \qquad (\sigma_Z\sigma_X\otimes I)
\end{align*}

For example, if Alice wants to send two classical bits $10$ to Bob, she will apply $\sigma_Z$ to her qubit before sending it to Bob. That would create the global state in Bob's possession  $\frac{1}{\sqrt{2}}\bigl(\ket{0_A0_B}-\ket{1_A1_B}\bigr)$, which he will learn after performing the Bell measurement.

\subsection{Teleportation}

Here again Alice and Bob had prepared together an entangled pair of qubits in the state \eqref{eq:EPRpair},
and then Alice took qubit $A$ and Bob took qubit $B$. Now, Alice has another qubit in the 
state
\[
\ket{\psi}=\alpha\ket{0}_a+\beta\ket{1}_a
\]
which she would like to send to Bob. (We will use $a$ (new) and $A$ (entangled with Bob) subscripts to distinguish the two qubits on Alice's side.) However, there is only a classical communications channel between Alice and Bob. Can Alice send her qubit to Bob by sending only classical bits of information? How many classical bits does she need to send?

To answer that question, consider the joint state of Alice's new qubit and the entangled pair:
\begin{align*}
\ket{\psi}\ket{\Psi} &= \bigl(\alpha\ket{0}_a+\beta\ket{1}_a\bigr)\frac{1}{\sqrt{2}}
\bigl(\ket{0_A}\ket{0_B}+\ket{1_A}\ket{1_B}\bigr)\\
&= \alpha\ket{0}_a\frac{1}{\sqrt{2}}\bigl(\ket{0_A0_B}+\ket{1_A1_B}\bigr)+
\beta\ket{1}_a\frac{1}{\sqrt{2}}\bigl(\ket{0_A0_B}+\ket{1_A1_B}\bigr)
\end{align*}
The following protocol, known as \textit{teleportation}  \cite{teleporting:bennett93}, results in Bob's qubit (member of the entangled pair) assuming the state $\ket{\psi}$:
\begin{enumerate}
	\item Alice first applies a {\tt CNOT} gate to her two qubits $\ket{x}_a\ket{x_A}$ obtaining $\ket{x_a}\ket{x_a\oplus x_A}$.
	The  3-qubit state then becomes
	\[
	\ket{\Phi} = \alpha\ket{0}_a\frac{1}{\sqrt{2}}\bigl(\ket{0_A0_B}+\ket{1_A1_B}\bigr)+
	\beta\ket{1}_a\frac{1}{\sqrt{2}}\bigl(\ket{1_A0_B}+\ket{0_A1_B}\bigr)
	\]
	\item Alice then applies a Hadamard transformation $H$ to her qubit $a$, and the joint state becomes
	\begin{align*}
	(H\otimes I\otimes I)\ket{\Phi}  & =
	\alpha \frac{1}{\sqrt{2}}(\ket{0}_A+\ket{1}_B)\frac{1}{\sqrt{2}}\bigl(\ket{0_A0_B}+\ket{1_A1_B}\bigr)\\
	&\,+
	\beta \frac{1}{\sqrt{2}}(\ket{0}_A-\ket{1}_B)\frac{1}{\sqrt{2}}\bigl(\ket{1_A0_B}+\ket{0_A1_B}\bigr)\\
	& = \frac{1}{2}\ket{00}_{aA}\bigl(\alpha\ket{0}_B+\beta\ket{1}_B\bigr)+
	\frac{1}{2}\ket{01}_{aA}\bigl( \alpha\ket{1}_B+\beta\ket{0}_B\bigr)\\
	& \,+\frac{1}{2}\ket{10}_{aA}\bigl( \alpha\ket{0}_B-\beta\ket{1}_B\bigr)+
	\frac{1}{2}\ket{11}_{aA}\bigl( \alpha\ket{1}_B-\beta\ket{0}_B\bigr)
	\end{align*}
	Observe the following:
	\begin{enumerate}
		\item The four states in the above sum are orthogonal.
		\item For each of the four basis states on Alice's side, we have a corresponding state on Bob's side that can be obtained from $\ket{\psi}$
		by a unitary action:
		\begin{align*}
		\alpha\ket{0}_B+\beta\ket{1}_B & = I\ket{\psi}   \\
		\alpha\ket{1}_B+\beta\ket{0}_B & = \sigma_X\ket{\psi}\\
		\alpha\ket{0}_B-\beta\ket{1}_B & = \sigma_Z\ket{\psi}\\
		\alpha\ket{1}_B-\beta\ket{0}_B & = \sigma_Z\sigma_X\ket{\psi}
		\end{align*}
	\end{enumerate}
	\item
	Alice performs a joint measurement of her two qubits in the computational basis. Her pair of qubits will collapse to one of the basis states and Bob's qubit will assume its corresponding state. After the measurement, Alice knows which state she is left with and thus which state Bob's qubit is in.  Bob can turn that state to $\ket{\psi}$ by applying the appropriate unitary operator. Whether that operator should be $I$, or  $\sigma_X$ or $\sigma_Z$ or $\sigma_Z\sigma_X$ can be communicated to him by Alice with 2 bits of classical information.  They have to have agreed on how to label the four operators.
	Observe that there is only one copy of state $\ket{\psi}$ at the end of the protocol, the one that Bob has. Alice's  2-qubit state collapsed to a basis state after her measurement. Therefore, teleportation is not cloning.
\end{enumerate}

\subsection{The CHSH Game\label{sec:CHSH}}
The CHSH game demonstrates how two players Alice and Bob, who cannot communicate with each other once the game starts, can benefit from shared entanglement much more than from shared classical randomness in winning the game \cite{CHSH:CleveHTW04}. It is rooted in a paper by Clauser, Horne, Shimony, and Holt, hence the name \cite{CHSH:clauserHSH69}.

In the CHSH game, Alice is given a binary input $x\in\{0,1\}$ and Bob is given a binary input $y\in\{0,1\}$ by a referee who guarantees that each combination of the inputs is equally likely.  Upon receiving the input, Alice generates her output $a$ and Bob his output $b$.  They send the outputs to  the referee who declares them the winners if $x \cdot y = a ~\text{\tt xor}~ b$. In other words, if $x=y=1$, Alice and Bob win if their outputs are different, and in all other cases, they win if their outputs are identical. Alice and Bob are allowed to agree on a strategy in advance, and to share random bits or entangled qubits, but once the game starts, they cannot communicate. Is there any advantage to be had from sharing entangled qubits?


A classical strategy to maximize the winning probability is that Alice and Bob send to the referee $a=b=0$ regardless of which 
input values they receive. With this strategy, they loose only when $x$ and $y$ are both $1$, and thus win the game with probability $0.75$.
 It is straightforward to check that this is the best strategy among the 16 deterministic strategies (ways to map four possible inputs to four possible outputs). Any shared classical randomness would essentially randomize among the 16  deterministic strategies, and thus cannot beat the best. The question then becomes if Alice and Bob can benefit from a shared EPR pair. The answer to this question is yes, and we next describe a strategy with the winning probability of about $0.85$.
 
\begin{figure}[hbt]
	\begin{center}
	\includegraphics[scale=0.99]{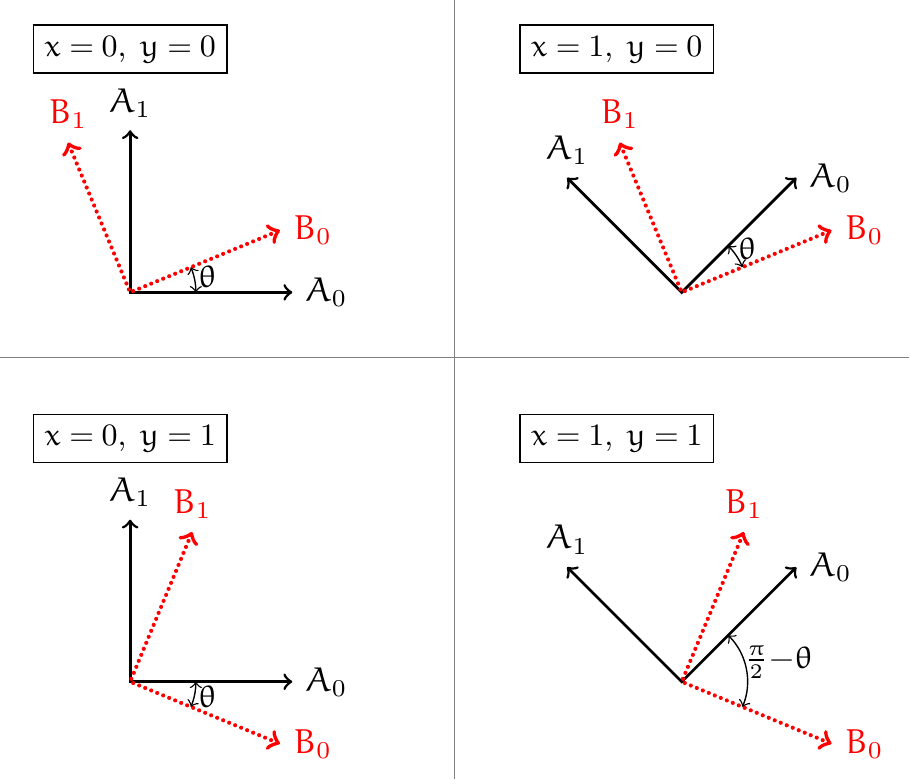}
	\end{center}
	\caption{Choice of measurement bases that Alice and Bob make based of the inputs $x$ and $y$. If Alice measures $A_i$, she  outpus $a=i$, and if Bob measures $B_i$, he outputs $b=i$. The angle $\theta$ is chosen to be $\pi/8$.}
	\label{fig:chshAB}
\end{figure}

 Consider a strategy where Alice and Bob share an entangled pair of qubits in the state $\ket{\varphi_{AB}}$ given in \eqref{eq:EPRpair}. Upon receiving the input, each player measures his/her qubit in one of the two possible bases depending on whether the input is $0$ or $1$. They then generate their outputs according to the result of the measurement. Alice's basis choices are the computational basis for input $x=0$, and the Hadamard basis for input $x=1$. Bob's two bases are as Alice's for the identical inputs just rotated 
 by $\pi/8$ and $-3\pi/8$, and he makes his choices based on his input $y$. Thus, there are four possible combinations of Alice/Bob measurement bases corresponding to the four different input pairs $x$ and $y$, as shown in Fig.~\ref{fig:chshAB}.
 %
In order to find the winning probability of this strategy, we next prove a general result about local measurements of entangled qubits.

Recall that Alice and Bob share an EPR pair in the state $\ket{\varphi_{AB}} $ given by \eqref{eq:EPRpair}.  Suppose that 
Alice measures her qubit in the basis $\bigl\{\ket{A_0},\ket{A_1}\bigr\}$ and Bob measurews his qubit in the basis $\bigl\{\ket{B_0},\ket{B_1}\bigr\}$, where $\ket{A_0}$ and $\ket{B_0}$ can be expressed in the computational basis as follows:
\[
|A_0\rangle = \cos\alpha |0\rangle + \sin \alpha |1\rangle ~~ \text{and} ~~
|B_0\rangle = \cos\beta |0\rangle + \sin \beta |1\rangle 
\]
Note that Bob's basis can be obtained from Alice's by rotation, where the rotation angle is $\alpha-\beta$. When Alice's measurement result is $i\in\{0,1\}$, she outputs $a=i$, and when Bob's measurement result is $j\in\{0,1\}$, he outputs is $b=j$. We next show that Alice and Bob will have identical outputs wp $\cos^2\theta$.

Since Alice and Bob perform their measurements locally, the measurement  on the shared EPR pair $\ket{\varphi_{AB}}$ is effectively performed  in the Kronceker product basis $\ops{A_i}\otimes\ops{B_j}$, $i,j\in\{0,1\}$.
 It follows from the definition of quantum measurement and simple geometry (see Fig.~\ref{fig:chshAB}) that 
\[ 
P(a=b=0)=P(a=b=1)=\bra{\varphi_{AB}}(\ops{A_0}\otimes\ops{B_0})\ket{\varphi_{AB}}.
\] 
Furthermore, we have
\begin{align*}
P(a=b) & = 2\cdot
\bra{\varphi_{AB}}(\ops{A_0}\otimes\ops{B_0})\ket{\varphi_{AB}}\\ &= \frac{2}{\sqrt{2}} \bigl(\ip{0}{A_0}\bra{A_0}\otimes \ip{0}{B_0}\bra{B_0}+\ip{0}{A_0}\bra{A_0}\otimes \ip{0}{B_0}\bra{B_0}\bigr)\ket{\varphi_{AB}}
\\ &= \cos^2\alpha\cos^2\beta + 2\cos\alpha\cos\beta\sin\alpha\sin\beta+\sin^2\alpha\sin^2\beta
\\ &= (\cos\alpha\cos\beta+\sin\alpha\sin\beta)^2=\cos^2 (\alpha-\beta)
\end{align*}
Therefore, when Bob's basis can be obtained from Alice's by the angle $\alpha-\beta$ rotation, we have
\begin{equation}
P(a=b) = \cos^2(\alpha-\beta) ~~ \text{and} ~~ P(a\ne b) = \sin^2(\alpha-\beta).
\label{eq:LocM}
\end{equation}

We are now ready to derive the probability that Alice and Bob win the CHSH game. Observe that 1) the angle between  Alice's and Bob's  measurement bases is $3\pi/8$ when $x=y=1$, and $\pi/8$ for all other input combinations (see in Fig.~\ref{fig:chshAB}), and 2) Alice and Bob win if they generate different outputs $a\ne b$ for inputs $x=y=1$, and identical outputs for all other input combinations. Therefore, by \eqref{eq:LocM}, we have
\begin{align*}
P(\text{Alice \& Bob win}) & =P(xy=0)P(a+b=0|xy=0)|+P(xy=1)P(a+b=1|xy=1)\\
& =  \frac{3}{4}P(a=b|xy=0)|+ \frac{1}{4}P(a\ne b|xy=1) ]\\
& = \frac{3}{4}\cdot\cos^2(\pi/8)+\frac{1}{4}\cdot \sin^2(3\pi/8)\\
& = \cos^2(\pi/8) = (2+\sqrt{2})/4 \gtrsim 0.853.
\end{align*}

\newpage
The significance of the CHSH game and similar tools is that they show that there is a limit to what can be done with classical (possibly hidden) randomness. If experiments involving shared entanglement show that this limit can be beaten (as they have), then there must be  some ``spooky action at a distance'' like the one reflected in \eqref{eq:LocM}. And that is where the weirdness and the power of quantum computing reside. 

\section{Conclusions}
This primer started with addressing the following three essential questions: How is quantum information represented? How is quantum information processed? How is classical information extracted from quantum states? 
The paper next focused on the fundamentals of quantum information theory. This material is provided to address the most common interests of this journal's readers. At the end, the paper presented the basics of the power of entanglement, which is, almost universally, considered to be the most fascinating feature of quantum information science. 

This paper presented the very basics of quantum information science in the most concise way the author could master. Its purpose was neither to survey this vast field nor speculate about its promising future, but rather to supply a bare minimum of materiel required to enter the field without necessarily taking an introductory class or reading a textbook. 
%
Entering the field of quantum information science means different things to different people, and ranges from starting to read research papers to joining multi-disciplinary teams in building quantum devices, as an expert in a related field. This paper strives to provide such diverse groups of scientists and engineers with what they need to know in order to start their endeavors and proceed to the next level.


\section*{Acknowledgment}
The author would like to thank Andrea da Conturbia, Esmaeil Karimi, and an anonymous reviewer for their comments on an earlier version of this paper.

\newpage

\bibliographystyle{IEEEtran}
\bibliography{QBref}
 \end{document}